\documentclass[aps,prd,twocolumn,showpacs,notitlepage,eqsecnum,
superscriptaddress]{revtex4-1}

\usepackage[centertags]{amsmath}
\usepackage{amssymb}
\usepackage{latexsym}
\usepackage{enumerate}
\usepackage{graphicx}
\usepackage{mathrsfs}
\usepackage{hyperref}
\usepackage{stmaryrd}
\usepackage{import}
\usepackage{tensor}
\usepackage{color}
\usepackage{bm}
\usepackage{multirow}
\usepackage{breakurl}
\usepackage{float}
\usepackage{placeins}

\newcommand{\bi}{\begin{itemize}}
\newcommand{\ei}{\end{itemize}}
\newcommand{\be}{\begin{equation}}
\newcommand{\ee}{\end{equation}}

\renewcommand{\l}{\left(}
\renewcommand{\r}{\right)}
\renewcommand{\a}{\alpha}
\renewcommand{\b}{\beta}
\newcommand{\g}{\gamma}
\newcommand{\G}{\Gamma}
\renewcommand{\d}{\delta}
\newcommand{\D}{\Delta}
\newcommand{\e}{\epsilon}
\newcommand{\ka}{\kappa}

\newcommand{\la}{\lambda}
\renewcommand{\O}{\Omega}
\renewcommand{\o}{\omega}
\renewcommand{\th}{\theta}
\newcommand{\Th}{\Theta}
\newcommand{\q}{\quad}
\newcommand{\qq}{\qquad}

\newcommand{\Si}{\Sigma}
\newcommand{\vp}{\varphi}

\newcommand{\pa}{\partial}

\allowdisplaybreaks

\begin{document}

\title{High-order post-Newtonian expansion of the generalized redshift invariant for eccentric-orbit, equatorial 
extreme-mass-ratio inspirals with a spinning primary}

\author{Christopher Munna}
\affiliation{Department of Physics and Astronomy, University of North Carolina, Chapel Hill, North Carolina 27599, USA}
\affiliation{MIT Kavli Institute, Massachusetts Institute of Technology, Cambridge, MA 02139, USA }

\begin{abstract}
We derive new terms in the post-Newtonian (PN) expansion of the generalized redshift invariant $\langle u^t \rangle_\tau$ 
for a small body in eccentric, equatorial orbit about a massive Kerr black hole.  The series is computed analytically using the 
Teukolsky formalism for first-order black hole perturbation theory (BHPT), along with the CCK method for metric 
reconstruction using the Hertz potential in ingoing radiation gauge.  Modal contributions with small values of $l$ are derived 
via the semi-analytic solution of Mano-Suzuki-Takasugi (MST), while the remaining values of $l$ to infinity are determined via 
direct expansion of the Teukolsky equation.  Each PN order is calculated as a series in eccentricity $e$ but kept exact in the 
primary black hole's spin parameter $a$.  In total, the PN terms are expanded to $e^{16}$ through 6PN relative order, and 
separately to $e^{10}$ through 8PN relative order.   Upon grouping eccentricity coefficients by spin dependence, we find that 
many resulting component terms can be simplified to closed-form functions of eccentricity, in close analogy to corresponding
terms derived previously in the Schwarzschild limit.  We use numerical calculations to compare 
convergence of the full series to its Schwarzschild counterpart and discuss implications for gravitational wave analysis.
\end{abstract}


\maketitle

\section{Introduction}
\label{sec:intro}

The radiative dynamics of binary black hole inspirals with an extreme mass ratio (that is, extreme-mass-ratio 
inspirals, or EMRIs) continues to be an active area of research.  Theoretical models must be able to predict the entire
trajectories of these inspirals to within a fraction of a radian over their lifetimes to produce accurate waveform 
templates for the coming space-based gravitational wave detector, LISA \cite{HindFlan08}.  

Over the last several years, we have sought to advance knowledge of EMRI motion and radiation through 
high-order post-Newtonian (PN) approximations to first-order black hole perturbation theory (BHPT) \cite{Munn20c}.  
This effort began with the case of eccentric-orbit inspirals on a Schwarzschild background, using the 
Mano-Suzuki-Takasugi (MST) solutions to the Regge-Wheeler-Zerilli (RWZ) equations \cite{ReggWhee57, Zeri70, 
ManoSuzuTaka96a,ManoSuzuTaka96b}.  With an implementation of this formalism in \textsc{Mathematica}, we 
were able to derive several notable features of the orbital evolution.  We first determined the (energy and angular
momentum) fluxes at infinity to 19PN, with each term expanded in eccentricity to $e^{10}$, as well as to 10PN and 
$e^{20}$ \cite{MunnEvan19a,MunnETC20,MunnEvan20a,Munn20}.  This effort was then extended to the horizon 
fluxes, which were computed to 18PN (relative order)/$e^{10}$ and 10PN/$e^{20}$ \cite{Munn20c, MunnEvanFors23}. 
 
It later proved possible to use this code to derive PN series for local gauge-invariant corrections to the conservative motion, 
including the redshift invariant (expanded to 10PN/$e^{20}$ \cite{MunnEvan22a}) and the spin-precession invariant 
(expanded to 9PN/$e^{16}$ \cite{MunnEvan22b}).  By pushing these series to such high orders, we enable greater 
access to the strong field regime and also inform the effective-one-body formalism (EOB), which can accurately 
describe binary dynamics across vast regions of parameter space \cite{BaraDamoSago10, LetiBlanWhit12, 
BiniDamo14c, BiniDamoGera15,Leti15,HoppKavaOtte16, KavaETC17, BiniDamoGera18, BiniDamoGera19, 
BiniDamoGera20a, BiniDamoGera20b}. 

As most astrophysical black holes have nonzero spin, it is necessary to understand the corresponding dynamics for 
EMRIs with a central Kerr black hole.  Therefore, we have more recently begun translating these techniques to the related 
Teukolsky formalism for perturbations about a Kerr background \cite{Teuk73}, which yields a similar (but more 
complicated) set of MST solutions for the relevant master functions \cite{ManoSuzuTaka96b, SasaTago03}.  As an 
intermediate step to fully generic inspirals, we have first restricted our efforts to the case of equatorial orbits, finding 
series for the fluxes at infinity to 8PN and $e^{20}$ \cite{MunnETC23}.  Along the way, we discovered many of the 
eccentricity series could be manipulated into exact (closed-form) functions.  The energy and angular momentum
absorbed by the central black hole has been found to a similar level, and those results will be published in a future 
paper \cite{MunnETC23b}.

As in the Schwarzschild case, we are now equipped to analyze the conservative sector of the first-order motion, 
which is known to contribute at the first post-adiabatic order in the full evolution and will thus be necessary for data 
analysis and parameter estimation with LISA \cite{HindFlan08}.  The most well-known quantity characterizing the
conservative sector is the redshift invariant $u^t$, which was first defined for circular Schwarzschild orbits and 
derived using the full PN theory to 3PN order by Detweiler \cite{Detw08} (see \cite{Blan14} for a review of PN 
theory).  As the significance of the redshift invariant in encoding the conservative dynamics became more widely 
appreciated, researchers began to derive deeper PN expansions in the small-mass-ratio limit using BHPT 
\cite{BiniDamo13, BiniDamo14a}, and this process was eventually carried out to 21.5PN \cite{KavaOtteWard15}.  

The extension to eccentric Schwarzschild orbits was first described by Barack and Sago \cite{BaraSago11}, who 
defined the so called generalized redshift invariant $\langle u^t \rangle_\tau$ as the proper-time average of $u^t$ 
over one radial libration.  The generalized redshift invariant was later derived to 3PN order in \cite{AkcaETC15} using 
the full PN theory and then subsequently higher order for EMRIs using BHPT \cite{BiniDamoGera15, 
HoppKavaOtte16, BiniDamoGera20a, BiniDamoGera20b}.  The most recent development was an expansion to 
10PN and $e^{20}$, with many PN terms found to yield closed-form functions of eccentricity \cite{MunnEvan22a}.  
That work also presented a set of curious connections between the (conservative-sector) redshift expansion and the 
(dissipative-sector) energy flux expansion, in that the two series share identical leading logarithm terms (see 
\cite{MunnEvan19a,MunnEvan20a,MunnEvan22a} for additional details). 

Because of its added difficulty, the redshift invariant for EMRIs with a Kerr primary has seen less progress, being first 
computed for circular equatorial EMRIs in 2012 \cite{ShahFrieKeid12}.  A BHPT-PN expansion was derived several years 
later to 8.5PN order \cite{BiniDamoGera15b}, with each PN term expanded in spin to $a^4$.  An 8.5PN series 
remaining exact in spin was then found in the work \cite{KavaOtteWard16}.  The eccentric, equatorial case was 
calculated numerically as part of a larger metric reconstruction effort in \cite{VandShah15}.  Then, the corresponding 
BHPT-PN expansion was derived in the small-$e$, small-$a$ limit to 8.5PN/$\mathcal{O}(e^2)$/$\mathcal{O}(a^2)$ 
in \cite{BiniDamoGera16c} and then to 8.5PN/$\mathcal{O}(e^4)$/$\mathcal{O}(a^2)$ in \cite{BiniGera19a}.
The work \cite{BiniGera19a} also produced a low-order derivation of the redshift within the full PN theory for
spinning bodies, using that result to confirm the first few terms in their BHPT-PN calculation.  

The present effort now seeks to extend this calculation beyond the nearly circular, nearly Schwarzschild regime by 
deriving results that are exact in $a$ and high order in $e$.  Specifically, we show series to 6PN and $e^{16}$ 
and to 8PN and $e^{10}$, both remaining exact in $a$.  To the author's knowledge, this is the first expansion of the 
redshift for eccentric orbits on a Kerr background with terms exact in $a$.  We assess the convergence of this 
series by comparing to numerical calculations for combinations from the sets $p \in \{10, 20\}, e \in \{1/10, 1/5\}, 
a \in \{1/4, 1/2, 9/10\}$ for semi-latus rectum $p$.  We find that convergence weakens with increasing $a$ and $e$, but that
the full expansion is accurate to better than one part in $10^4$ for most of these orbits.  This calculation will serve as a final 
intermediate step on the path to generic (eccentric/inclined) inspirals on a Kerr background, which has not yet been 
computed analytically or numerically (though the numerical infrastructure for generic orbits does now exist 
\cite{Vand17}).  

Calculation of the redshift invariant requires the local regularized metric perturbation, which can be found via the
CCK (Chrzanowski, Cohen, Kegeles) metric reconstruction procedure \cite{Chrz75,CoheKege74,KegeCohe79, 
Wald78}.  We use the MST-Teukolsky solutions to form the Hertz potential and then apply a sequence of linear 
operations to produce components of the perturbed metric at the location of the smaller body in ingoing radiation 
gauge.  We find that, as in the Schwarzschild case, the leading PN order of each $l$ mode is constant in $l$, 
necessitating PN series for all $l$.  This difficulty is resolved using a PN ansatz solution for large $l$ that is general in 
$l$ \cite{BiniDamo13, KavaOtteWard15, KavaOtteWard16}.  Thus, we use the MST solutions for small $l \ge 2$ and 
the ansatz for large $l$, along with a separate metric completion procedure 
for $l=0$ and $l=1$.  This general-$l$ ansatz solution can also be expanded about $l=\infty$ to determine the 
divergent behavior of the summation and then regularize each $l$-mode of the full solution.  In total the process
is relatively similar to that for eccentric Schwarzschild EMRIs, though the introduction of the spin parameter $a$
and loss of spherical symmetry add several technical hurdles and greatly increase the computational complexity.

The structure of this paper is as follows.  In Sec.~\ref{sec:MSTreview} we briefly outline the problem setup for 
first-order BHPT on a Kerr background and the Teukolsky-MST formalism in the PN limit.  We then discuss how to
apply the (PN-expanded) MST solutions to the CCK procedure for metric reconstruction using the Hertz potential, as 
the local metric perturbation is the primary constituent of the redshift invariant.  Sec.~\ref{sec:genLexps} details 
the derivation of metric perturbation expansions for general $l$, with emphasis on the unique theoretical and 
computational challenges contained therein.  In Sec.~\ref{sec:l01andReg} we briefly review the metric completion
piece and our chosen regularization scheme for the redshift invariant.  Sec.~\ref{sec:ut} then details the 
explicit expansion results to 6PN/$e^{16}$ and 8PN/$e^{10}$, which are also posted in multiple online repositories
\cite{BHPTK18,UNCGrav22}.  Multiple new closed-form expressions are found, and the structure of the expansion's
spin-dependence is discussed, as well as its convergence against numerical data.  Sec.~\ref{sec:ConsConc} concludes with 
summary and outlook.

Throughout this paper we apply the metric signature $(-+++)$ and primarily choose units such that $c = G = 1$, 
though we frequently retain powers of $\eta = 1/c$ to track PN order.  Our notation for the Teukolsky and MST 
formalisms follows that found in \cite{SasaTago03, KavaOtteWard16}. 

\section{Review of the Teukolsky and MST formalisms}
\label{sec:MSTreview}

We briefly review the background and setup for first-order perturbations about a Kerr background caused by a 
small mass in equatorial orbit.  At each step we seek expressions that are suitable for expansion in the PN limit,
whether through the direct parameter $\eta = 1/c$ or through a measure of orbital separation such as $1/p$ for
semi-latus rectum $p$.  These methods are more extensively detailed in \cite{MunnETC23} and \cite{Munn20}, 
based on earlier Kerr work in \cite{BiniDamoGera15b,BiniDamoGera16c,KavaOtteWard16,BiniGera19a} and 
Schwarzschild work in \cite{BiniDamo13,BiniDamo14a,KavaOtteWard15,BiniDamoGera16a,HoppKavaOtte16}.

\subsection{Bound equatorial orbits on a Kerr background}
\label{sec:orbits}

At lowest order, the secondary is treated as a point mass $\mu$ in bound geodesic orbit about a Kerr black hole of 
mass $M$ with $\varepsilon = \mu/M \ll 1$.  The line element in Boyer-Lindquist coordinates 
$x^{\mu} = \{t,r,\theta, \vp \}$ is
\begin{align}
ds^2 &= - \left( 1 - \frac{2 M r}{\Si} \right) dt^2 - \frac{4 M a r \sin^2 \th}{\Si} dt d\vp + \frac{\Si}{\D} dr^2 \notag \\&
+ \Si d\th^2
+ \left( r^2 + a^2 + \frac{ 2 M a^2 r \sin^2 \th}{\Si} \right) \sin^2 \th d\vp^2 ,
\end{align}
where $\Si=r^2+a^2\cos^2\theta$, $\D =r^2-2Mr+a^2$, and $a$ is the spin of the primary.  

We now restrict the orbit to the equatorial plane, which leads to the following equations of motion:
\begin{align}
\label{eqn:geoEqns}
\left(r^2 \frac{dt}{d\tau}\right) &= \left(a \mathcal{L} - a^2 \mathcal{E} \right) + \frac{r^2 + a^2}{\D} 
\left( \mathcal{E} (r^2 + a^2) - a \mathcal{L} \right)  ,   \notag \\
\left(r^2 \frac{dr}{d\tau}\right)^2 & = [\mathcal{E} (r^2 + a^2) - a \mathcal{L}]^2 
- \D[(a \mathcal{E} - \mathcal{L})^2 + r^2]  ,   \notag  \\
\left(r^2 \frac{d\vp}{d\tau}\right) &= \mathcal{L} - a \mathcal{E} +\frac{a}{\D} \left(\mathcal{E} (r^2 + a^2) 
- a \mathcal{L}\right)  .
\end{align}
Here $\mathcal{E}$ is the (conserved) specific energy and $\mathcal{L}$ the specific angular momentum.  As
noted in \cite{MunnETC23}, these equations of motion are only dependent on the radial coordinate $r(\tau)$.  This 
implies that we do not have to invoke the use of Mino time $d\la = d\tau/\Si$ and can instead move immediately to 
the Darwin parameterization.
As is typical for Schwarzschild geodesics, we describe the motion in terms of the set $\{ \chi, p, e \}$ for relativistic
anomaly $\chi$, semi-latus rectum $p$, and eccentricity $e$ \cite{Darw59,CutlKennPois94,BaraSago10}, with
\begin{align}
r_p \l \chi \r = \frac{pM}{1+ e \cos \chi} .
\end{align}
One radial libration occurs with each $2 \pi$ advance in $\chi$. Then, defining $\tilde{a} = a/M$ and 
$\hat{\mathcal{L}}=\mathcal{L}/M - \tilde{a} \mathcal{E}$, we find the following relations \cite{BiniGeraJant16}
\begin{align}
\mathcal{E}^2 &=  \frac{(1-e^2)^2}{p^3} \hat{\mathcal{L}}^2 + 1 - \frac{1 - e^2}{p}  , \notag \\
\mathcal{E} &= - \frac{ p - 3 - e^2 }{2 \tilde{a} p} \hat{\mathcal{L}} - \frac{\tilde{a}^2 - p}{2\tilde{a} \hat{ \mathcal{L}}}  .
\end{align}
These equations can be solved exactly for $\mathcal{E}(p,\tilde{a},e)$ and $\hat{\mathcal{L}}(p,\tilde{a},e)$, though the 
results are lengthy (they are given in \cite{MunnETC23}).  However, we note that $1/p$ is a standard PN parameter, in 
which $\mathcal{E}$ and $\hat{\mathcal{L}}$ can be easily expanded.  To simplify the process somewhat, we define 
$v = 1/\sqrt{p}$ and compute series about $v = 0$.  Then, just as in the Schwarzschild case, we expand each PN order in 
$e$ to prepare for the eventual source integration.  On the other hand, we make no approximations with respect to $a$, 
leaving each step in the expansion process exact in that parameter.

Applying these definitions and relations into the equations \eqref{eqn:geoEqns} leads to the following ODEs for
the coordinates:
\begin{widetext}
\begin{align}
\frac{d \tau}{d\chi} &= \frac{M}{v^3 (1+e \cos (\chi ))^2 \sqrt{1 + \hat{\mathcal{L}}^2 v^4 (e^2 - 2 e \cos \chi - 3)} } 
\notag \\
\frac{dt}{d\chi} &= \left(\frac{d\tau}{d\chi}\right)  \frac{\mathcal{E} + \mathcal{E} \tilde{a}^2 v^4 (1+ e \cos \chi)^2 
- 2 \tilde{a} v^6 \hat{\mathcal{L}} (1+e\cos\chi)^3}{1 - 2 v^2 (1+ e\cos\chi) + \tilde{a}^2 v^4 (1+ e\cos\chi)^2 }  \notag \\ 
\frac{d\vp}{d\chi} &= \frac{v [\mathcal{L} - 2 v^2 \hat{\mathcal{L}} (1+ e \cos \chi)]}{[1 - 2 v^2 (1+ e\cos\chi) 
+ \tilde{a}^2 v^4 (1+ e\cos\chi)^2]\sqrt{1 + \hat{\mathcal{L}}^2 v^4 (e^2 - 2 e \cos \chi - 3)}  } .
\end{align}
These equations are then readily PN expanded in $v$ and $e$, and the result is trivially
integrated to yield expansions for $t(\chi)$ and $\vp(\chi)$.  They can also be solved exactly using elliptic integrals
\cite{FujiHiki09}.  Then, the radial period is given by $T_r = t(2 \pi)$, the radial frequency by $\O_r = 2 \pi/T_r$,
and the azimuthal frequency by $\vp(2 \pi)/T_r$.

\subsection{The Teukolsky master equations}
\label{sec:TDmasterEq}

Bound motion acts as a periodic source for the first-order gravitational perturbations.  On a Kerr background these 
can encoded by a set of Teukolsky master functions ${}_{s}R_{l m\o}$ in radiation gauge \cite{Teuk73, SasaTago03} 
and associated spin-weighted spheroidal harmonics ${}_{s}S_{l m\o}$.  We will focus on the functions with 
spin-weight $s=-2$, which are governed by the equations,
\begin{align}
 \D^2 \frac{d}{dr}\left( \frac{1}{\D}\frac{d {}_{-2}R_{l m\o}}{dr} \right) + \bigg[\frac{K^2+4i(r-M)K}{\D}
 - 8i\o r - {}_s \la_{lm\o} \bigg] {}_{-2}R_{lm\o} = T_{lm\o}  ,  \\
\bigg[ \frac{1}{\sin\theta} \frac{d }{d\th} \left(\sin\th \frac{d}{d\th} \right) -a^2\o^2\sin^2\theta  
- \frac{(m - 2 \cos\th)^2}{\sin^2\th} + 4 a \o \cos\th - 2 +2 ma\o+{}_{-2}\la_{lm\o} \bigg] {}_{-2}S_{l m \o}=0  .
\end{align}
\end{widetext}
Here $K = (r^2+a^2) \o - ma$, ${}_{-2}\la_{lm\o}$ is the spin-weighted spheroidal eigenvalue, and $T_{lm\o}$ is the 
decomposition of the (Newman-Penrose projection of the) stress-energy tensor.  The spheroidal harmonic also has 
the normalization condition
\begin{align}
\int_0^{\pi} |_{-2}S_{l m \o}|^2 \sin\theta d\theta=1 .
\end{align}

Ultimately, these functions encode the first-order perturbation through composition of the quantity $\psi_4$, a certain 
Newman-Penrose projection of the Weyl tensor given by
\be
\varrho^{-4} \psi_4= \sum_{l m} \int e^{-i\o t+i m \vp} {}_{-2}S_{l m \o}(\th) R_{l m\o}(r)  d\o ,
\ee
where $\varrho^{-1} = - (r - i a \cos \th)$.  For more detail on the motivation and derivation behind these equations, see 
\cite{Teuk73, SasaTago03}.  A deeper discussion is also given in \cite{MunnETC23}.

Because the source motion is biperiodic, the Fourier integral collapses into a Fourier sum over the frequencies
$\o = \o_{mn} = m \O_\vp + n \O_r$.  Then, the homogeneous form of the equation yields two independent solutions: 
${}_{-2}R_{lmn}^{-} = {}_{-2}R_{lmn}^{\rm in}$, with causal (ingoing wave) behavior at the horizon, and 
${}_{-2}R_{lmn}^{+} = {}_{-2}R_{lmn}^{\rm up}$ with causal 
(outgoing wave) behavior at infinity.  Both can be derived as infinite sums of hypergeometric functions using the MST 
formalism \cite{ManoSuzuTaka96b,SasaTago03}, which we briefly review below. 

The spin-weighted spheroidal harmonics can be expanded in the spheroidicity $a\o$ (a 1.5PN quantity) by inserting 
a PN ansatz, applying standard boundary and normalization conditions, then solving a system of equations 
\cite{KavaOtteWard16}.  However, we expand them (along with their first and second $\th$-derivatives) simply by 
using the spheroidal harmonic package of the black hole perturbation toolkit \cite{BHPTK18}.  The coefficient on 
each power of $(a\o)$ is given by a finite sum of spin-weighted spherical harmonics, which then takes an analytic 
value for specific choices of $s,l,m,$ and $\th$.  See \cite{MunnETC23} for additional details.

\begin{widetext}
\subsection{The MST homogeneous solutions and the source integration}

The MST solution for ${}_{-2}R_{lmn}^+$ can be expressed \cite{SasaTago03, KavaOtteWard16} as
\begin{align}
\label{eqn:RupMST}
{}_{-2}R^{+}_{lmn} = e^{iz} \frac{z^{\nu + i(\e+\tau)/2}}{(z - \e \ka)^{-2 + i(\e + \tau)/2}}  \sum_{j=-\infty}^\infty 
a_j^{\nu} (2 i z)^{j} \frac{\G(j + \nu - 1 - i \e) \G(\nu + 3 + i \e)}{\G(j + \nu + 3 + i \e) \G(\nu -1 - i \e) }  
U(j+\nu - 1 - i \e, 2j + 2\nu + 2, -2 i z), 
\end{align}
where $U$ is the irregular confluent hypergeometric function, $\e = 2 M \o \eta^3$, $z = (r-r_-) \o \eta$, 
$r_\pm = GM (1 \pm \ka)\eta^2$, $\ka = \sqrt{1-a^2}$, and $\eta = 1/c$.  The parameter $\nu$ is the renormalized 
angular momentum, an eigenvalue chosen to make the series coefficients $a_j$ (not to be confused with the spin 
parameter) converge as $j \rightarrow \pm\infty$.  Both $\nu$ and $a_j$ are determined through continued 
fraction expansion \cite{ManoSuzuTaka96b,SasaTago03}, which leads to series in $\e$ for both.  The rest of the
formula can then be expanded in both $\e$ and $z$, leading to a composite PN series in $\eta$ (which by definition
is a 0.5PN expansion parameter) \cite{KavaOtteWard16, MunnETC23}.  

Similarly, the solution for ${}_{-2}R_{lmn}^-$ can be written as
\begin{align}
\label{eqn:RinMST}
{}_{-2}R^{-}_{lmn} = e^{-iz + i \ka \e} \left( \frac{\e \ka}{z} \right)^{i\tau +s} 
\left(1- \frac{\e \ka}{z}\right)^{-s - i(\e + \tau)/2} 
\sum_{j=-\infty}^{\infty} a_j^{\nu}  {}_2F_1(j+\nu+1 - i\tau,-j - \nu - i \tau, 3 - i\e - i\tau,1-\frac{z}{\e \ka}) .
\end{align}
Here, ${}_2F_1$ is the ordinary hypergeometric function, and the parameters $\nu$ and $a_j$ are the same as
in Eq.~\eqref{eqn:RupMST}.

The process of expanding these homogeneous solutions by collecting on powers of $\eta$ is fully described in 
\cite{MunnETC23}, based on the methods initially presented in \cite{BiniDamoGera15b,BiniDamoGera16c, 
KavaOtteWard16}, as well as \cite{Munn20}.  From a computational perspective, the most important consideration is 
the fact that the homogeneous solutions
as written in \eqref{eqn:RupMST} and \eqref{eqn:RinMST} contain many cumbersome $z$-independent factors 
that greatly complicate the expansion \cite{KavaOtteWard16,Munn20,MunnETC23}.  These factors will eventually 
cancel through division by the Wronskian and can thus be omitted from the start.  One useful choice of normalization
is given in \cite{MunnETC23}, and the process is operationally similar to the Schwarzschild case, described in
\cite{Munn20}.  When computing the fluxes, the solutions must eventually be rescaled to produce the 
proper asymptotic behavior; however, in the metric reconstruction procedure this problem is avoided entirely, as 
all choices of normalization lead to the same result.

Once appropriately simplified, the homogeneous solutions can be used to complete the source integration 
\cite{SasaTago03,MunnETC23}:
\begin{align}
\label{eqn:ZlmnIntChi}
{}_{-2}Z_{lmn}^{\pm} &= \frac{1}{W_{lmn} T_r } \int_{0}^{T_r} \left[ (A_{nn0} + A_{\bar{m}n0} 
+ A_{\bar{m}\bar{m}0})R_{lmn}^{\mp} - (A_{\bar{m}n1} + A_{\bar{m}\bar{m}1} ) \frac{d R_{lmn}^{\mp}}{dr} 
+ A_{\bar{m}\bar{m}2} \frac{d^2 R_{lmn}^{\mp}}{dr^2} \right] e^{i \o t - i m \vp(t)} dt 
\end{align}
The Wronskian $W_{lmn}$ is given by 
\be
W_{lmn} = \frac{1}{\D} \bigg[ \frac{d {}_{-2}R_{lmn}^+}{dr} {}_{-2}R_{lmn}^- 
-  \frac{d {}_{-2}R_{lmn}^-}{dr} {}_{-2}R_{lmn}^+ \bigg] ,
\ee
and source $A$ functions (deriving from $T_{lm\o}$) are defined in Sasaki and Tagoshi \cite{SasaTago03} for generic 
orbits.  The equatorial limits can be found in \cite{MunnETC23}.

\subsection{The Hertz potential in ingoing radiation gauge}
\label{sec:Hertz}

The first-order generalized redshift invariant is a quantity that depends upon the behavior of the regularized metric 
perturbation along the particle's worldline.  The global metric perturbation in radiation gauge can be derived using the CCK 
procedure \cite{Chrz75,CoheKege74,KegeCohe79, Wald78}.  In short, an intermediate Hertz potential $\Psi$ is first 
constructed from products of the normalization coefficients ${}_{-2}Z^{\pm}_{lmn}$, the homogeneous solutions 
${}_{-2} R^{\pm}_{lmn}$, and spin-weighted spheroidal harmonics ${}_{-2} S_{lmn}$.  The Hertz potential is then transformed 
through a sequence of linear operations to yield the metric perturbation.  

The $s=-2$ solutions are most easily adapted to the use of ingoing radiation gauge, whose Hertz potential in 
vacuum is a solution of the homogeneous $s=-2$ Teukolsky equation \cite{LousWhit02,KeidETC10}.  
Thus, it can be written in the generic form
\be
\label{eqn:HertzPot}
\Psi^\pm= \frac{1}{\sqrt{2 \pi}} \sum_{lmn} \Psi_{lmn}^\pm \, {}_{-2} R_{lmn}^\pm(r) \, 
{}_{-2} S_{lmn}^\pm \,e^{im\vp - i \o t}  ,
\ee
for some undetermined coefficients $\Psi_{lmn}^\pm$.  In general, it is also governed by the following fourth-order 
inhomogeneous partial differential equations \cite{LousWhit02,KeidETC10}:
\begin{align}
\frac{1}{2} (\boldsymbol{D})^4 \bar{\Psi} &= \psi_0, \notag \\
\frac{1}{8} [ \tilde{\mathcal{L}}^4 \bar{\Psi} - 12 M \pa_t \Psi ] &= \varrho^{-4} \psi_4, 
\end{align}
where
\begin{align}
\boldsymbol{D} &= \frac{r^2 + a^2}{\D} \pa_t + \pa_r + \frac{a}{\D} \pa_\vp = l^\mu \pa_\mu ,  \\
\tilde{\mathcal{L}}^4 &= \tilde{\mathcal{L}}_1 \tilde{\mathcal{L}}_0 \tilde{\mathcal{L}}_{-1} \tilde{\mathcal{L}}_{-2}  \\
\tilde{\mathcal{L}}_q &= -\pa_\th - q \cot \th + i \csc \th \pa_\vp + i a \sin \th \pa_t ,
\end{align}
and the overbar denotes complex conjugation.

The angular equation can be used to identify the coefficients $\Psi_{lmn}^\pm$.  First, we must note that the complex 
conjugate can be expressed as 
\begin{align}
\bar{\Psi}^\pm &= \frac{1}{\sqrt{2 \pi}} \sum_{lmn} \bar{\Psi}_{lmn}^\pm \, {}_{-2} \bar{R}_{lmn}^\pm(r) \, 
{}_{-2} \bar{S}_{lmn}^\pm \,e^{-im\vp + i \o t}  \notag \\
 &= \frac{1}{\sqrt{2 \pi}} \sum_{lmn} \bar{\Psi}_{lmn}^\pm \, {}_{-2} R_{l-m-n}^\pm(r) \, 
(-1)^m {}_{2} S_{l-m-n}^\pm \,e^{-im\vp + i \o t}  \notag \\
&= \frac{1}{\sqrt{2 \pi}} \sum_{lmn} \bar{\Psi}_{l-m-n}^\pm \, {}_{-2} R_{lmn}^\pm(r) \, 
(-1)^m {}_{2} S_{lmn}^\pm \,e^{im\vp - i \o t}  
\end{align}
where we used the identity ${}_s \bar{S}_{lmn} = (-1)^{m+s} {}_{-s} S_{l-m-n}$ and took $(m,n) \rightarrow (-m,-n)$ 
in the sum.  Then, the $\tilde{\mathcal{L}}^4$ operator is simplified using the Teukolsky-Starobinsky identity from 
Chandrasekhar \cite{Chan83}, or
\be
\tilde{\mathcal{L}}^4 \, {}_{2} S_{lmn} = F  \,({}_{-2} S_{lmn})
\ee
for
\begin{align}
F^2 &= ({}_{-2} \la_{lmn})^2 ({}_{-2}\la_{lmn} + 2)^2 + 8 a\o (m-a\o)({}_{-2}\la_{lmn}) [5 {}_{-2} \la_{lmn}+6] 
+48 a^2 \o^2 [2 ({}_{-2} \la_{lmn}) + 3 (m - a \o )^2] .
\end{align}

With these steps the amplitudes are found to satisfy the relation
\begin{align}
\frac{1}{8} [ F (-1)^m \bar{\Psi}_{l-m-n} + 12 M (i\o) \Psi_{lmn} ] = {}_{-2}Z_{lmn}^\pm,
\end{align} 
This can be inverted to isolate $\Psi_{lmn}$ by taking a linear combination of $Z_{lmn}$ and 
$Z_{l-m-n}$.  We get
\begin{align}
\label{eqn:Hertzcoef}
\Psi_{lmn} = 8 \frac{ (-1)^m F Z_{l-m-n} - 12 M i\o Z_{lmn}}{F^2 + 144 M^2 \o^2} = 
 8 \left[\frac{ (-1)^{l+m} F - 12 M i\o }{F^2 + 144 M^2 \o^2}\right] Z_{lmn},
\end{align}
where the last step applied the identity $Z_{l-m-n} = (-1)^l Z_{lmn}$.  Thus, the coefficients are relatively easy to 
retrieve from the $s=-2$ homogeneous solutions and normalization constants.  

\subsection{The metric perturbation in ingoing radiation gauge}
\label{sec:metPerExps}

With the Hertz potential computed, the metric perturbations in ingoing radiation gauge follow as \cite{MerlETC16}
\begin{align}
\label{eqn:pmunu}
p_{\mu\nu}&= - \bigg\{\ell_\mu \ell_\nu(\boldsymbol{\delta} + \bar{\a} + 3\b - \tau)(\boldsymbol{\delta}+ 4\b+3\tau)
+m_\mu m_\nu(\boldsymbol{D} - \varrho)(\boldsymbol{D} + 3 \varrho)  \notag \\&
- \ell_{(\mu}m_{\nu)}\left[(\boldsymbol{\d} - 2\bar{\a}+2 \beta -\tau)(\boldsymbol{D} + 3 \varrho)
+(\boldsymbol{D}+ \bar{\varrho} - \varrho)(\boldsymbol{\delta} + 4\b+ 3\tau)\right]\bigg\}\Psi+\text{c.c.} .
\end{align}
The various operators, tetrad components, and spin coefficients are defined in the Newman-Penrose formalism
\cite{NewmPenr62}.  Explicitly, we have
\begin{align}
\ell^\alpha &= \frac{1}{\Delta}(r^2+a^2,\Delta,0,a), \notag \\
m^\alpha &=-\frac{\bar{\varrho}}{\sqrt{2}}(i a \sin \theta,0,1,\frac{i}{\sin \theta}),   \notag \\
\boldsymbol{\delta} &= m^\mu \pa_\mu = \frac{ia \sin \th}{\sqrt{2} (r + i a\cos\th)} \pa_t 
+ \frac{1}{\sqrt{2} (r + i a\cos\th)} \pa_\th +  \frac{i}{\sqrt{2}  (r + i a\cos\th) \sin \th} \pa_\vp , \notag \\
\varrho &=\frac{-1}{r-i a \cos \theta},
\qquad \tau=\frac{-i a \sin \theta}{\sqrt{2} \Sigma} ,  
\qquad \beta = \frac{\cot \theta}{2\sqrt{2}(r + i a \cos \theta)} , \notag \\
\alpha &= \frac{i a \sin \theta }{\sqrt{2}(r-i a \cos \theta)^2} - \frac{\cot \theta}{2\sqrt{2} (r + i a \cos \theta)}  .
\end{align}
Note that the IRG metric perturbation satisfies the conditions $\ell^\mu p_{\mu \nu} = 0$ and 
$g^{\mu \nu}_{\rm Kerr} p_{\mu \nu} = 0$, where $g^{\mu \nu}_{\rm Kerr}$ is the (inverse) background metric.  

Once these tetrad terms, along with the modal form of the Hertz potential, are inserted into \eqref{eqn:pmunu}, the 
result is an unwieldy combination of ${}_{-2}R_{lmn}, \pa_r {}_{-2}R_{lmn}, \pa_r^2 {}_{-2}R_{lmn}$ and 
${}_{-2}S_{lmn}, \pa_\th {}_{-2}S_{lmn}, \pa_\th^2 {}_{-2}S_{lmn}$ multiplying factors involving $r, \o, m,$ and $\th$. 
The full expression simplifies in the equatorial plane, but the result remains too large to display here.  Nevertheless, 
once the metric perturbation is derived in analytic form, the process of completing its PN expansion
is straightforward, if cumbersome.

\section{General-$l$ expansions}
\label{sec:genLexps}

The MST formalism described in the previous section provides mode functions for specific $l$.  In the dissipative
sector \cite{Munn20, MunnETC23}, PN expansions of the relevant observables (e.g., the fluxes) possess leading
behavior that increases with $l$.  Thus, an expansion to any particular desired PN order requires calculation of only 
finite values of $l$.  Unfortunately, this phenomenon does not recur in the conservative dynamics, as the leading PN 
order of the local metric perturbations $p_{\mu \nu}^{l}(\chi)$ is constant in $l$.  As a result, we must compute 
expansions for all values of $l$ to determine the full metric perturbation $p_{\mu \nu}(\chi)$, which is prohibitively 
difficult using the MST approach.  In the Schwarzschild case, it proved possible to use a PN ansatz solution in the 
RWZ equation to obtain expansions that remained general in $l$, which could then be iterated through the rest of the 
process and summed over all values of $l$ when necessary \cite{BiniDamo13, BiniDamo14a, KavaOtteWard15, 
MunnEvan22a}.  It turns out that a similar approach can be completed in the Kerr problem \cite{KavaOtteWard16}, 
though with several added difficulties over the Schwarzschild background.  We detail the full procedure below.

\subsection{The homogeneous solutions and normalization constants}

As in the Schwarzschild case \cite{BiniDamo13,KavaOtteWard15}, we start by introducing a PN ansatz for the 
homogeneous solutions of the Teukolsky equation.  Following \cite{KavaOtteWard16}, we choose
\begin{align}
{}_sR_{lmn}^- &=\left(\frac{\bar{\e}}{\bar{z}} \eta^2\right)^{-\nu + s} (1 + A_1 \eta + A_2 \eta^2 + \cdots 
+ A_{2l} \eta^{2l} + \mathcal{O}(\eta^{2l+1}) ) ,
\notag \\
{}_sR_{lmn}^+ &= (\bar{z} \eta)^{-\nu-1-s} (1+B_1 \eta+ B_2 \eta^2 + \cdots + B_{2l} \eta^{2l} 
+ \mathcal{O}(\eta^{2l+1})),
\end{align}
where the $A_i$ and $B_i$ are functions of $(\bar{z}, \bar{\e}, l, m, a)$, $\bar{z}=r\o$, and $\bar{\e}=2GM\o$.  Once 
$\nu$ is found using the continued fraction method for general $l$, these expressions are plugged into the ($s=-2$)
homogeneous Teukolsky equation,
\begin{align}  
\D^3 \frac{d}{dr}\left( \frac{1}{\D}\frac{dR_{lmn}}{dr} \right) + [K^2+4i(r-M)K - (8i\o r+ {}_{-2}\la_{lmn} ) \D ] R_{lmn}= 0 ,
\end{align}
and solved order-by-order.  Note that we multiplied the original Teukolsky equation by $\D$ to simplify appearances
of $\eta$.  Unfortunately, the ansatz does not fully apply the boundary conditions, which is why it breaks down after 
some $l$-dependent PN order \cite{BiniDamo13,BiniDamo14a, KavaOtteWard15, KavaOtteWard16}.
If a target PN order $P$ is set, the ansatz will be useless for $l \lesssim P$.  Thus, those values of $l$ must be 
determined separately with the MST formalism.

Proceeding in this way, we obtain a general-$l$ PN expansion for $\nu$, which begins
\begin{align}
\nu &= l+\frac{24+13 l+28 l^2+30 l^3+15 l^4}{6 l+10 l^2-20 l^3-40 l^4-16 l^5} \e^2 + 
\frac{(108+18 l+17 l^2+3 l^3+14 l^4+15 l^5+5 l^6) }{l^2 (1+l)^2 (6+l-29 l^2-6 l^3+20 l^4+8 l^5)} ma \,\, \e^3  
+\mathcal{O}(\e^4)
\end{align}
and obtain the general-$l$ expansions for the mode functions, which likewise begin
\begin{align}
{}_{-2}R_{lmn}^{+,\rm ser} &= (\bar{z} \eta)^{l-1} {}_{-2}R_{lmn}^{+} = 1-\frac{2 i \bar{z}}{l} \eta+\frac{\bar{\e} l (2l -1) 
\left(l^2-1+2 i a m\right) +\left(l^2 - 7l -8 \right) \bar{z}^3 }{2 l (1+l) (2l - 1) \bar{z}} \eta ^2 
-\frac{i (l-3) \bar{z}^3}{(l-1) l (2l-1)}   \eta^3 +  \notag \\&
\frac{1}{8 \bar{z}^2} \bigg[\frac{\bar{\e}^2 \left(2 l^4 - (a^2 - 4) l^3-9 a^2 m^2+l^2 (8iam - 2) 
+l (12 i a m - 4-a^2 (m^2-1) ) \right)}{(1+l) (3+2 l)} 
+\frac{\left(32-17 l+l^2\right) \bar{z}^6}{l (2l-3) \left(1-3 l+2 l^2\right)}\notag \\&
+\frac{2 \bar{\e} \left(14 l^4+l^5+16 i a m+16 l (1-i a m)+l^3 (-7+2 i a m)+l^2 (-4+12 i a m)\right) \bar{z}^3}{l^3 \left(-1+l
+2 l^2\right)}\bigg] \eta ^4  + \mathcal{O}(\eta^5) , \notag \\
{}_{-2}R_{lmn}^{-,\rm ser}  &=  \left(\frac{\bar{\e}}{\bar{z}} \eta^2\right)^{l+2} {}_{-2}R_{lmn}^{-} = 
1+\frac{2 i \bar{z}}{1+l} \eta - \frac{1}{2\bar{z}}\left[\bar{\e} \left(2+l+\frac{2 i a m}{l}\right)
+\frac{(9+l) \bar{z}^3}{3+5 l+2 l^2}\right] \eta^2 -\frac{i (4+l) \bar{z}^3}{(2+l) (3+5 l+2 l^2)} \eta^3  \notag \\&
+\bigg[ \frac{\bar{\e}^2 (l (1+l) (2+l)  (a^2+2 l -2)+4 i a (1+l) (2l -1) m+a^2 (l-8) m^2 )}{l (2l-1)}
+\frac{(50+l (19+l)) \bar{z}^6}{(1+l) (2+l) (3+2 l) (5+2 l)}  \notag \\&
+\frac{2 \bar{\e} (l (1+l) (-48+l (-43+(-10+l) l))+2 i a (-21+l (-17+(-3+l) l)) m) \bar{z}^3}{l (1+l)^3 (3+2 l)}\bigg] \eta ^4  
+ \mathcal{O}(\eta^5)
\end{align}
Here, we defined $R^{\rm ser}_{lmn}$ as the normalized PN series that begin at $\mathcal{O}(1)$, which are more 
convenient to manipulate at each step of the calculation than the original series with $l$-dependent PN orders.  
Eventually, all $l$-dependent powers of $\eta$ will cancel in the metric perturbation due to their corresponding 
presence in the Wronskian. 

Once expanded, the general-$l$ homogeneous solutions can be evaluated along the geodesic orbit to compute the 
Wronskian and prepare for the source integration (at which point the series will be defined in terms of $v$ and $e$, as in the
specific-$l$ case).  However, the source terms themselves carry one additional complication: expansions of the 
spin-weighted spheroidal harmonics become large and unwieldy for general $l$ and $m$, which significantly slows 
the procedure.  We avoid this problem by leaving ${}_{-2}S_{lmn}$ and its derivatives as unevaluated parameters
until the final step of metric reconstruction.  Once the metric perturbation is expanded, we handle products of the 
spheroidal harmonics together in the sum over $m$.

\subsection{Sums of spin-weighted spheroidal harmonics over $m$}

The general-$l$ expansions are carried through the source integration and then used to construct the metric 
perturbation \eqref{eqn:pmunu} just as in the specific-$l$ case.  Computationally, the source integration is completed
in one step, and then the expansions for $Z_{lmn}^+, R_{lmn}^+, S_{lmn},$ and the Fourier kernel $e^{im\vp-i\o t}$ 
are directly included in the formula for the metric perturbation.  Each $n$ is calculated individually, as only finite $n$ 
modes are required to reach any particular order in $e$, and then the set is summed at the end.  We also split up the 
calculation over the nine spheroidal harmonic products ($S_{lmn}, \pa_\th S_{lmn}, \pa_{\th\th} S_{lmn}$ within 
the normalization constant $Z_{lmn}^+$, multiplied against the same three terms within the rest of the metric perturbation 
\eqref{eqn:pmunu}) that are left unevaluated until the end.  Even with this extensive segregation of terms, the PN expansion 
for each individual component serves as the computational bottleneck for this procedure.  In particular, the
$(n=1, S_{lmn} \times S_{lmn})$ part of each metric perturbation requires about 8 days and 10GB of memory to reach either
6PN/$e^{16}$ or 8PN/$e^{10}$ on the UNC supercomputing cluster Longleaf.  (Fortunately, it is trivial to parallelize over 
the various combinations of $n$ and $S_{lmn}$.)

Once all the components are calculated, we are left with the task of summing the mode expressions over $m$ to obtain 
$p_{\mu \nu}^l$.  This process is nontrivial, as each $m$ mode contains products of spin-weighted spheroidal 
harmonics (still unexpanded), and the sum must be taken from $-l$ to $l$ for general 
$l$.  In the Schwarzschild-RWZ problem, we faced a similar obstacle, having to complete sums of the form,
\begin{align}
\sum_{m=-l}^{l} m^{N} Y_{lm}(\pi/2,0)^2  , \qq \qq
\sum_{m=-l}^{l} m^{N} \pa_{\theta}Y_{lm}(\pi/2,0)^2 ,
\end{align}
where $Y_{lm}$ is the standard scalar spherical harmonic,
\be
Y_{lm}(\th,\vp) = \sqrt{\frac{(2 l+1)(l-m)!}{4 \pi (l+m)!}} P_l^m(\cos\th) e^{im\vp} ,
\ee
and $N$ is any positive integer.  In that case the first sum can be derived using a special case of the spherical harmonic 
addition theorem \cite{MunnEvan22a}:
\be
\label{eqn:addtheorem}
\sum_{m=-l}^l e^{im\vp}  Y_{lm}(\pi/2,0)^2 = \left(\frac{2 l + 1}{4 \pi} \right) P_l(\cos\vp)  ,
\ee
with the result for each value of $N$ corresponding to a term in the Taylor expansion of this formula about $\vp=0$.  The 
second summation could then be derived from derivatives of the spherical harmonic addition theorem, or from
a hypergeometric generating function \cite{NakaSagoSasa03}. 

In the Kerr-Teukolsky formalism, we now encounter sums like 
\begin{align}
\label{eqn:slmSums}
T_{00}^0 &=\sum_{m=-l}^{l} m^{N} {}_{-2}S_{lm}(\pi/2)^2 , \qq \qq \qq \qq \qq ~~
T_{00}^1 = \sum_{m=-l}^{l} (-1)^{l+m} m^{N} {}_{-2}S_{lm}(\pi/2)^2  , \notag \\
T_{01}^0 &=\sum_{m=-l}^{l} m^{N}{}_{-2}S_{lm}(\pi/2)  (\pa_\th {}_{-2}S_{lm}(\pi/2)) , \qq \q ~
T_{01}^1 = \sum_{m=-l}^{l} (-1)^{l+m} m^{N} {}_{-2}S_{lm}(\pi/2) (\pa_\th {}_{-2}S_{lm}(\pi/2)) ,  \notag \\
T_{11}^0 &=\sum_{m=-l}^{l} m^{N} (\pa_\th {}_{-2}S_{lm}(\pi/2))^2 , \qq \qq \qq \qq \q
T_{11}^1 = \sum_{m=-l}^{l} (-1)^{l+m} m^{N} (\pa_\th {}_{-2}S_{lm}(\pi/2))^2  , 
\end{align}
and so on for sums $T_{02}^{0/1}, T_{12}^{0/1}, T_{22}^{0/1}$.  Note that we have suppressed the $n$ index for
convenience and that the factors of $(-1)^{l+m}$ can be traced back to the Hertz potential coefficients in 
\eqref{eqn:Hertzcoef}.  Unfortunately, the spin-weighted spheroidal harmonics do not have a known addition 
theorem, and it is unlikely that any comparable formula like it can be derived.  In fact, ${}_{-2}S_{lm}(\pi/2)$ 
does not even have a known analytic form for specific values of $l$ and $m$.

Nevertheless, progress can be made by PN expanding $S_{lm}$ and its derivatives, now for general-$lm$, yielding
series like \cite{KavaOtteWard16, BHPTK18}
\begin{align}
\label{eqn:slmExp}
{}_s S_{lm}(\th) &= {}_s Y_{lm}(\th)+\frac{s}{\sqrt{1+2 l}} \left(\frac{\sqrt{l^2-m^2} \sqrt{l^2-s^2}}{l^2 \sqrt{2 l -1}}
{}_s Y_{(l-1)m}(\th)-\frac{\sqrt{(l+1)^2-m^2} \sqrt{(l+1)^2-s^2} }{(1+l)^2 \sqrt{3+2 l}} {}_s Y_{(l+1)m}(\th) \right) a \o  
  \notag \\&
+ \frac{1}{2} \bigg[-\frac{\sqrt{l^2-m^2} \sqrt{1-2 l+l^2-m^2} (l-2 s^2 ) \sqrt{l^2-s^2} \sqrt{1-2 l+l^2-s^2}
}{(1-2 l)^2 (-1+l) l^2 \sqrt{-3+2 l} \sqrt{1+2 l}}  {}_s Y_{(l-2)m}(\th)  + {}_s Y_{(l-1)m}(\th) \times \notag \\&
\frac{2 m s \sqrt{(l^2-m^2)(l^2-s^2)}  (l^2-2 s^2 ) }{l^4 \sqrt{4 l^2 - 1} (l^2-1)}  
+s^2 \left( \frac{\left(l^2-m^2\right) \left(l^2-s^2\right)}{l^4 \left(1 - 4 l^2\right)}  
-\frac{ [(l+1)^2-m^2][(l+1)^2-s^2]}{(1+l)^4 \left(3+8 l+4 l^2\right)}\right) {}_s Y_{lm}(\th) \notag \\&
-\frac{2 m \sqrt{1+2l+l^2-m^2} s \left(1+2 l+l^2-2 s^2\right) \sqrt{1+2 l+l^2-s^2} }{l (1+l)^4 (2+l)
\sqrt{1+2 l} \sqrt{3+2 l}}  {}_s Y_{(l+1)m}(\th)  + \notag \\&
\frac{\sqrt{[(l+1)^2-m^2][(l+2)^2-m^2]}\sqrt{[(l+1)^2-s^2][(l+2)^2-s^2]} \left(1+l+2 s^2\right)
}{(1+l)^2 (2+l) \sqrt{1+2 l} (3+2 l)^2 \sqrt{5+2 l}} {}_s Y_{(l+2)m}(\th) \bigg] (a \o)^2  +\cdots
\end{align}
Derivatives of ${}_s S_{lm}(\th)$ are applied directly to the terms in the series.  Thus, when products of these series
are taken, we will be left with sums over spin-weighted spherical harmonics, instead of spheroidal harmonics.  

Spin-weighted spherical harmonics are also unlikely to yield straightforward summation formulas, and the reference 
\cite{KavaOtteWard16} handled these sums using Mathematica's \texttt{FindSequenceFunction}.  However, it turns 
out that exact, if cumbersome, formulas can be derived analytically by using the spin-weighted spherical harmonic 
definition to transform back to scalar spherical harmonics.  Explicitly,
\begin{align}
{}_{s-1} Y_{lm} &= \frac{(\sin\th)^{-s}}{\sqrt{(l+s)(l-s+1)}} \left[\frac{\pa}{\pa \th}
 - \frac{i}{\sin\th} \frac{\pa}{\pa \vp} \right] ((\sin \th)^{s} {}_s Y_{lm}) .
\end{align}
Thus, $s=-2$ can be expressed in terms of $s=0$ as
\begin{align}
{}_{-2}Y_{lm} =  \sqrt{\frac{(l-2)!}{(l+2)!}} \bigg[ \pa_\th^2 Y_{lm} 
+ \frac{2 m - \cos\th}{\sin \th} \pa_\th Y_{lm} +  \frac{m^2-2m\cos\th}{(\sin \th)^2} Y_{lm} \bigg] .
\end{align}
Substitution of this relation and \eqref{eqn:slmExp} into the summation formulas \eqref{eqn:slmSums} would then 
yield products of scalar spherical harmonics alone, which are closely connected to the spherical harmonic addition
theorem.  However, the expression \eqref{eqn:slmExp} contains multiple values of the first harmonic number 
($l,l\pm 1, l \pm 2$, etc.), while the addition theorem \eqref{eqn:addtheorem} only explicitly covers products of terms 
with identical harmonic numbers.  This last problem can be resolved by using the well-known scalar spherical 
harmonic identity,
\begin{align}
Y_{(l+1)m} &=  \cos\th \sqrt{\frac{(2l+1)(2l+3)}{(l+1)^2-m^2}} Y_{lm} 
- \sqrt{\frac{(2l+3)(l^2-m^2)}{(2l-1)[(l+1)^2-m^2]}}  Y_{(l-1)m}  .
\end{align}
We also find it convenient to eliminate all derivative terms using identities of the following form:
\begin{align}
\sin\th \frac{d}{d \th} Y_{lm} &= l \cos\th \, Y_{lm} - \sqrt{\frac{(2l+1)(l^2-m^2)}{(2l-1)}}  Y_{(l-1)m}  , \notag \\
\sin\th \frac{d}{d \th} Y_{lm} &= - (l+1) \cos\th \, Y_{lm} + \sqrt{\frac{(2l+1)[(l+1)^2-m^2]}{(2l+3)}}  Y_{(l+1)m}  .
\end{align}

In total we make a sequence of identity transformations until all terms are of the form $Y_{lm}$ and $Y_{(l-1)m}$.  
Then, the remaining products will involve only $m^N Y_{lm}^2$ and $m^N Y_{(l-1)m}^2$, which are trivial to execute 
using the addition theorem.  In particular, we find that no square root terms appear in the final product and that the 
cross terms vanish, as $Y_{lm}(\pi/2)$ is 0 whenever $(l+m)$ is odd, meaning one of $Y_{lm}(\pi/2)$ and 
$Y_{(l-1)m}(\pi/2)$ is always 0.  Parity considerations also clarify how to account for factors of $(-1)^{m}$ --- 
these terms simplify to become overall factors of $(-1)^l$ or $(-1)^{l-1}$, as appropriate.  Then, the full term $(-1)^{l+m}$ 
simply contributes $(-1)^{2l} = 1$ or $(-1)^{2l-1} = -1$ to each component of the sum.

Thus, to summarize, we expand the spin-weighted spheroidal harmonics into series of spin-weighted spherical harmonics; 
then, we use the definition of each ${}_{-2}Y_{lm}$ to re-express it in terms of $Y_{lm}$.  Next, $\th$-derivatives and 
distant values of the first harmonic number are eliminated using identities, at which point the standard spherical 
harmonic addition theorem can be used to complete the summation.  Once the metric perturbation is summed over $m$, the 
general-$l$ expansion for $p_{\mu \nu}^l$ will be ready for use in the construction of the redshift. 

\section{Metric completion and regularization}
\label{sec:l01andReg}

With the procedure established for the specific-$l$ (MST) and the general-$l$ (ansatz) parts of the metric 
perturbation, we are left with two remaining considerations: the completion of the metric (monopole and dipole
terms) and regularization procedure.  We briefly cover those issues here.

\subsection{Non-radiative modes}
\label{sec:l0l1}

The Teukolsky formalism is only valid for the modes $l \ge s$.  Notably, it omits the corrections to the mass monopole
and dipole of the primary (informally referred to as the $l=0$ and $l=1$ modes), which must be derived separately.
The full completion part $p_{\mu\nu}^{\rm comp}$ was first derived in \cite{MerlETC16}.  It is given by 
\begin{align}
p^{{\rm comp}+}_{tt} &= \frac{2r}{\Sigma^2}[(r^2+3a^2\cos^2\theta)\delta M-2a\cos^2\theta\delta J] \,, \notag \\
p^{{\rm comp}+}_{rr} &= \frac{2r}{M\Delta^2}\{[M(r^2+3a^2\cos^2\theta)+a^2r\sin^2\theta]\delta M 
-a[r \sin^2\th + 2 M \cos^2\theta]\delta J\} \,,\notag\\ 
p^{{\rm comp}+}_{\theta\theta} &= - \frac{2a}{M}\cos^2\theta(a\delta M-\delta J)  \,,\notag\\ 
p^{{\rm comp}+}_{\phi\phi} &= - \frac{2a}{M\Sigma^2}\sin^2\theta\{a[\Sigma^2 +Mr\sin^2\theta(r^2
-a^2\cos^2\theta)]\delta M -(\Sigma^2+2Mr^3\sin^2\theta)\delta J\}  \,,\notag\\
p^{{\rm comp}+}_{t\phi}&= -\frac{2r}{\Sigma^2}\sin^2\theta[2a^3\cos^2\theta\delta M +(r^2-a^2\cos^2\theta)\delta J]
\,,
\end{align}
where $p_{\mu\nu}^{\rm comp} = p_{\mu\nu}^{{\rm comp}+} \Th [r-r_p(t)]$.  After a lengthy calculation, 
\cite{MerlETC16} confirmed the expected result that $\d M = \mu \mathcal{E}$ and $\d J = \mu \mathcal{L}$.

Because we are only interested in the local perturbation, we can restrict these expressions to the equatorial
plane, which simplifies the result to
\begin{align}
p^{{\rm comp}+}_{tt} &= \frac{2 \d M}{r}  \,, \notag \\
p^{{\rm comp}+}_{rr} &= \frac{2r^2}{M\Delta^2} [(M r + a^2)\delta M - a \delta J ] \,,\notag\\ 
p^{{\rm comp}+}_{\theta\theta}&= 0  \,,\notag\\ 
p^{{\rm comp}+}_{\phi\phi} &= - \frac{2a}{M r} [a (r +M)\delta M - (r + 2M)\delta J ]  \,,\notag\\
p^{{\rm comp}+}_{t\phi} &= - \frac{2}{r} \delta J \,. 
\end{align}

\end{widetext}

Actually, there is another non-radiative contribution termed $p^{\rm gauge}_{\mu\nu}$ that was discussed at 
length in \cite{BiniGera19d}.  However, this term is 0 for all $r>r_p$ and therefore does not affect the value of the 
redshift, which is calculated here in the limit $r \rightarrow r_p$ from above.  The authors of \cite{BiniGera19d} also 
note that the redshift combination $p_{\mu \nu}^R u^\mu u^\nu$ must be continuous across $r=r_p$, so that the 
gauge portion is not needed.  On the other hand, the spin-precession invariant \cite{DolaETC14a,AkcaDempDola17} 
is typically regularized through an upper-lower-limit averaging procedure, so it is likely that 
$p^{{\rm gauge}+}_{\mu\nu}$ will have to be included in that calculation.

\subsection{Mode-sum regularization}
\label{sec:regularization}

Thus far, the expressions given for the metric perturbation in ingoing radiation gauge have referred to the full
retarded field.  This field formally diverges at the location of the particle, a property that becomes apparent when
the $l$ modes are summed from $l=0$ to $l=\infty$.  Local gauge invariant quantities derived from these modes then
exhibit the same behavior.  Instead, from the full retarded field we must extract the so called regular field, which
defines the effective metric experienced by the smaller body.  

The regular part of the metric is derived through regularization, which can be achieved in a number of ways.  One
popular approach was given by Detweiler and Whiting \cite{DetwWhit03}, which chooses a particular split of 
regular and singular fields,
\be
p_{\mu \nu}(x) = p_{\mu \nu}^{S}(x) + p_{\mu \nu}^{R}(x) .
\ee
With this choice the singular field $p_{\mu \nu}^{S}(x)$ satisfies the same inhomogeneous field equation as 
$p_{\mu \nu}(x)$ but with different boundary conditions, while $p_{\mu \nu}^{R}$ then solves the homogeneous
field equations.  The orbiting particle then travels on the Kerr metric plus the perturbation $p_{\mu \nu}^{R}$.

Determination of the singular field is a difficult process \cite{HeffOtteWard12a}.  In first-order BHPT a common 
approach is the mode-sum regularization procedure \cite{Bara01,BaraOri03b}, which exploits the fact that the individual 
$l$-modes of the retarded metric perturbation are finite.  The part of the $l$-dependence that diverges in the 
infinite sum is subtracted off each $l$-mode, so that the full sum remains finite.  

The $l$-dependence of the singular field can be expressed as an expansion about $l=\infty$.  The metric 
perturbation itself will have a leading coefficient independent of $l$, which will obviously diverge in the sum over $l$.
Each derivative of the singular field with respect to any of the coordinates will increase the divergence by a power
of $l$.  In Lorenz gauge, it is known that the large-$l$ expansion can be manipulated into a form such that only
the divergent terms are needed to obtain the regular field --- the rest will vanish in a sum over all $l$.  Interestingly,
in a numerical calculation of finite $l$, the higher-order terms (regularization parameters) can improve convergence
\cite{HeffOtteWard12a}.  However, our analytic calculation will complete the infinite sum over all $l$, so only the 
divergent terms are needed.

The redshift invariant is directly proportional to the metric perturbation, so what remains is to determine the 
$l$-independent coefficient (the leading regularization parameter) in its expansion about $l=\infty$.  In fact, this can 
be done using our general-$l$ PN series.  This series is expanded in $l$, and the leading coefficient produces the 
singular field.  There is a subtlety involving our use of radiation gauge, instead of the more established Lorenz 
gauge, as the former is related to the latter by an irregular gauge transformation \cite{PounMerlBara14}.  However, it 
was noted by Detweiler \cite{Detw08} that the regularization scheme becomes gauge invariant when working with 
certain gauge invariant quantities, the redshift among them.  This $l$-expansion approach to regularization has 
already been used successfully to construct the PN series for redshift invariant of eccentric, equatorial (Kerr) EMRIs 
in \cite{BiniDamoGera16c, BiniGera19a}.

\section{The generalized redshift invariant}
\label{sec:ut}

\subsection{Background and implementation}

The generalized redshift invariant has the same definition and interpretation for eccentric, equatorial inspirals on 
a Kerr background as it does for eccentric inspirals on a Schwarzschild background.  Thus, the corresponding
discussion in our previous Schwarzschild work \cite{MunnEvan22a}, based on prior derivations in \cite{BaraSago11, 
AkcaETC15, HoppKavaOtte16}, is sufficient to understand the meaning and significance of the PN expansion 
presented here.  Nevertheless, we will recapitulate the development of the generalized redshift invariant 
here for the sake of completeness.

The redshift invariant $u^t$ was originally constructed for quasi-circular inspirals \cite{Detw05, Detw08}.  Note that
this quantity is precisely the inverse of the redshift itself, $z = 1/u^t$.   For eccentric orbits, Barack and Sago 
discovered that the proper-time average over a radial libration $\langle u^t \rangle_\tau$ provided the 
more appropriate gauge invariant measure of the conservative dynamics.  This average is equal to the 
coordinate-time period, $T_r$, divided by the proper-time period, $\mathcal{T}_r$.  To subleading order in the mass
ratio, this quotient is given by \cite{BaraSago11, AkcaETC15}
\be
 \left<u^t\right>_\tau = \frac{T_r}{\mathcal{T}_r + \D \mathcal{T}_r} =
 \frac{T_r}{\mathcal{T}_r} -  \D \mathcal{T}_r \frac{T_r}{\mathcal{T}_r^2} 
=  \left<u^t\right>_\tau^0 +  \left<u^t\right>_\tau^1  .
\ee
Note that the coordinate-time period $T_r$ is not corrected because the frequencies are held fixed from zeroth to 
first order (which is necessary for the gauge invariance of the redshift invariant \cite{BaraSago11}).  The leading term 
$\langle u^t \rangle_\tau^0$ is merely the value of the redshift invariant for geodesic orbits, which is trivial to 
calculate using the Darwin parameterization described in Sec.~\ref{sec:orbits}.  The second term, which incorporates
the effects of the first-order conservative self-force, can be shown to take the form \cite{BaraSago11, AkcaETC15}
\be
\D \mathcal{T}_r = - \mathcal{T}_r \left< \frac{1}{2} p_{\mu \nu}^R u^{\mu} u^{\nu}\right>_\tau .
\ee
This formula is the same for eccentric, equatorial orbits on both Schwarzschild and Kerr backgrounds.  Thus, the
correction to the generalized redshift invariant follows directly from the regularized metric perturbation, which we 
have detailed extensively in previous sections for the purpose of PN expansion.  

As mentioned before, this particular gauge-invariant quantity encodes important details of the conservative motion of 
the system.  The first-order conservative dynamics contribute at $\mathcal{O}(\varepsilon^0)$ in the cumulative 
EMRI phase, a level needed for the creation of accurate waveform templates in the LISA mission, making the 
redshift invariant especially valuable \cite{HindFlan08}.  In addition, there is an exact correspondence between the 
PN expansion of $\langle u^t \rangle_\tau^1$ and several important quantities in EOB theory.  For instance, the 
eccentric part of $\langle u^t \rangle_\tau^1$ can be used to derive the expansion of the $Q(1/r, p_r ;\nu)$ 
EOB potential, which governs the deviation from geodesic behavior in the EOB Hamiltonian \cite{Leti15, 
DamoJaraScha15, BiniDamoGera16a, HoppKavaOtte16,BiniDamoGera16c}.  The transformation between these 
quantities is outlined in \cite{Leti15}.  The circular spin-dependent part, meanwhile, is critical to the radial equatorial
potential $A(r,m_1,m_2,S_1,S_2)$ and the main spin-orbit coupling potential $G_S(r, m_1, m_2, S_1, S_2)$
\cite{BiniDamoGera16c}.  The eccentric spin-dependent part is expected to be more fruitful still, though the 
precise transcription scheme has not yet been elucidated \cite{BiniGera19a}.
 
The last remaining task is to implement the mode-sum regularization scheme in order to ensure a proper, convergent
sum over the $l$-modes of $\left<u^t\right>_\tau^1$.  We choose to regularize the final averaged product, which
is already in gauge invariant form:
\begin{align}
\left< \frac{1}{2} p_{\mu \nu}^R u^{\mu} u^{\nu}\right>_\tau &= 
\sum_{l=0}^\infty \left< \frac{1}{2} (p_{\mu \nu}^l - p_{\mu \nu}^{S,l}) u^{\mu} u^{\nu}\right>_\tau \notag\\
&=\sum_{l=0}^\infty \left<\frac{1}{2} p_{\mu\nu}^l u^{\mu} u^{\nu}\right>_\tau -\left< H_{[0]} \right>_\tau .
\end{align}
The singular field contribution is thus distilled down to an $l$-independent constant, equal to its leading behavior in a 
large-$l$ expansion, in accordance with the observations of the previous section. 

Analytic derivations of the singular part of the redshift invariant have been made \cite{BaraSago11, 
HeffOtteWard12a, HoppKavaOtte16}, but these have generally utilized a decomposition over spherical harmonics, 
while our $l$ modes derive from spin-weighted spheroidal harmonics.  For circular, equatorial orbits on a Kerr 
background, Kavanagh, Ottewill, Wardell derived the spin-weighted spheroidal form of $H_{[0]}$ to 13PN order 
\cite{KavaOtteWard16}.   We chose to use the general-$l$ ansatz expansion to extract the large-$l$ behavior directly, 
as has been done in \cite{BiniDamoGera16c, BiniGera19a}.  Our expanded $H_{[0]}$ is confirmed to match the 
analytic result of \cite{KavaOtteWard16} in the circular orbit limit.  The PN series for the orbital average of $H_{[0]}$
begins
\begin{align}
\left< H_{[0]} \right>_\tau &= (1-e^2) v^2 - \left[ (1-e^2)^2 - \frac{3}{4} (1-e^2)^{3/2} \right] v^4  \notag \\&
- 2 \tilde{a} [ (1-e^2)^2 - (1-e^2)^{3/2} ] v^5   \notag \\&
+ \bigg[ (1-e^2)^{3/2} \bigg( \frac{153}{64} + \frac{3}{4} \tilde{a}^2 +\frac{267 e^2}{128} \bigg)    \notag \\&
- (1-e^2) ( 3 + \tilde{a}^2 + e^2 )  \bigg] v^6 + \mathcal{O}(v^7)
\end{align}
This constant is then subtracted off at each value of $l$ from $l=0$ to $l=\infty$.  Note that this covers all three 
regimes of calculation: the metric completion piece ($l=0$ and $l=1$), the MST specific-$l$ solutions from $l=2$ to
$l=7$, and the general-$l$ ansatz solution from $l=8$ to $l=\infty$.  The form of the summands will involve products 
and quotients of polynomials in $l$, which are trivial to sum in \textsc{Mathematica}.  

For the simpler Schwarzschild problem, the same basic procedure was first implemented in \cite{BiniDamoGera16a}, 
where the (first-order BHPT) redshift invariant was expanded to 6.5PN and $e^2$ in eccentricity and to 4PN and 
$e^4$.  This was quickly extended by \cite{HoppKavaOtte16} to 4PN through $e^{10}$.  Later, 
\cite{BiniDamoGera20b} improved the eccentric knowledge to 9.5PN and $e^{8}$, as that level was needed to 
complete a novel transcription of the redshift invariant to the scattering angle for hyperbolic orbits, which can be used 
to compute the full post-Minkowskian dynamics to high order.  Finally, our previous work \cite{MunnEvan22a} brought 
the eccentric Schwarzschild expansion to 10PN and $e^{20}$.  Interestingly, by taking the expansions to such high 
order in $e$, we were able to find many PN terms which could be manipulated into either closed-form expressions or 
infinite series with known coefficients, following similar developments in the fluxes \cite{ForsEvanHopp16, 
MunnEvan19a, MunnETC20, MunnEvan20a}.  In fact, it was discovered that the entire leading logarithm series of 
the energy flux at infinity \cite{MunnEvan19a, MunnEvan20a} exactly reappears in the redshift invariant.  

This paper now extends many of those same advances to the more difficult Kerr problem, which has historically 
seen much less development.  The first expansion for eccentric, equatorial orbits was undertaken in 
\cite{BiniDamoGera16c}, finding a result to 8.5PN/$\mathcal{O}(e^2)$/$\mathcal{O}(a^2)$ in both a small-$e$
and small-$a$ limit.  This was later extended to 8.5PN/$\mathcal{O}(e^4)$/$\mathcal{O}(a^2)$ in \cite{BiniGera19a},
which also derived an expression to 3.5PN and $\mathcal{O}(a^2)$ using the full PN theory (i.e., for arbitrary mass
ratio).  We use the techniques and simplifications discussed in earlier sections of this paper to enhance these calculations 
greatly to the level of 6PN/$e^{16}$ and 8PN/$e^{10}$, all while remaining exact in $a$.  We then apply many of the 
techniques developed in the Schwarzschild case to extract closed-form eccentricity functions for the certain 
spin-dependent parts of the series.  Note that in the expressions presented below, we redefine $a$ to be 
dimensionless (i.e., $a \rightarrow a/M = \tilde{a}$) for simplicity.

\begin{widetext}
\subsection{PN expansion of the redshift invariant}

For eccentric, equatorial orbits, the first-order BHPT part of the generalized redshift invariant is found to take
the following form, mirroring its circular-orbit limit \cite{KavaOtteWard16, MunnEvan22a}:
\begin{align}
 \left<u^t\right>_\tau^1 
=& \, \, \left(\frac{\mu}{M}\right) \frac{1}{p} \bigg[ \mathcal{U}_0 + \frac{ \mathcal{U}_1}{p} 
+ \frac{ \mathcal{U}_{3/2}}{p^{3/2}} 
+ \frac{ \mathcal{U}_2}{p^2} + \frac{ \mathcal{U}_{5/2}}{p^{5/2}} + \frac{\mathcal{U}_{3} }{p^3}
+ \frac{ \mathcal{U}_{7/2}}{p^{7/2}} 
+ \Big( \mathcal{U}_{4} + \mathcal{U}_{4L} \log p \Big) \frac{1}{p^4}
+ \frac{ \mathcal{U}_{9/2}}{p^{9/2}}    \notag \\& \qq
+ \Big( \mathcal{U}_{5} + \mathcal{U}_{5L} \log p \Big) \frac{1}{p^5}
+ \Big( \mathcal{U}_{11/2} + \mathcal{U}_{11/2L} \log p \Big) \frac{1}{p^{11/2}}
+ \Big( \mathcal{U}_{6} + \mathcal{U}_{6L} \log p \Big) \frac{1}{p^6}   \notag \\& \qq
+ \Big( \mathcal{U}_{13/2} + \mathcal{U}_{13/2L} \log p \Big) \frac{1}{p^{13/2}}  
+ \Big( \mathcal{U}_{7} + \mathcal{U}_{7L} \log p 
+ \mathcal{U}_{7L2} \log^2 p \Big) \frac{1}{p^7}  \notag \\& \qq
+ \frac{\mathcal{U}_{15/2}}{p^{15/2}}  
+ \Big( \mathcal{U}_{8} + \mathcal{U}_{8L} \log p 
+ \mathcal{U}_{8L2} \log^2 p \Big) \frac{1}{p^8}  
+ \cdots \bigg]  .
\end{align}
Note that we restore use of the parameter $1/p = v^2$ in this section, as it is more commonly used in the literature.  This 
expansion exhibits two key differences from its Schwarzschild counterpart.  The first is that while in the Schwarzschild 
limit each term $\mathcal{U}_i$ was a function of eccentricity alone, now each $\mathcal{U}_i=\mathcal{U}_i(a,e)$ is 
a function of both eccentricity $e$ and spin $a$.  In many cases we will be able to extract their exact dependence 
on both parameters, though often our results will be Taylor expanded in $e$.  The second is the presence 
of half-integer terms starting at the 1.5PN level.  These terms are purely spin-dependent, as the first half-integer PN 
term in the non-spinning case appears at 5.5PN.   

In this work, we present only the spin-dependent coefficients, as the $a=0$ limit was discussed at length in 
\cite{MunnEvan22a}.  To support this effort, we define an additional layer of specification at each order,
\begin{align}
\mathcal{U}_i(a,e) = \mathcal{U}_i(e)^{\rm Sch} + \sum_{k=0} a^k \mathcal{U}_i(a,e)^{Sk} ,
\end{align}
such that the first term $\mathcal{U}_i(e)^{\rm Sch}$ corresponds to the Schwarzschild limit, and the superscript $k$ 
describes the power of $a$ attached to the remaining functions.  All terms through
6PN were found through $e^{16}$ (if not exactly), while the orders 6.5PN-8PN were found to $e^{10}$.  Some of the 
functions at higher orders are too lengthy to display in their entirety.  These are truncated after a few eccentricity
coefficients; however, we add a Greek letter (e.g., $\a_{16}$) to the end of such functions to remind the reader of the 
extent of the series.  The full results are made available in electronic form on the black hole perturbation toolkit 
\cite{BHPTK18}, as well as the UNC gravity repository \cite{UNCGrav22}.

We begin with the functions through 3.5PN order, which all yield closed-form expressions.  The first two terms are 
entirely spin-independent, so we list the spin-dependent enhancement functions from 1.5PN-3.5PN:
\begin{align}
\mathcal{U}_{3/2}^{S1} &= - 2 (1-e^2)^2 + 5 (1-e^2)^{3/2} , \notag \\
\mathcal{U}_2^{S2} &= (1-e^2)^2 - 2 (1-e^2)^{3/2}, \notag \\
\mathcal{U}_{5/2}^{S1} &= (1-e^2)^2 (-20 + 8 e^2) + (1-e^2)^{3/2} (38 + 5 e^2), \notag \\
\mathcal{U}_3^{S2} &= (1-e^2)^2 (13 - e^2) - (1-e^2)^2 (27 + 29 e^2), \notag \\
\mathcal{U}_{7/2}^{S1} &= (1-e^2)^2 \left(-\frac{87}{2}-\frac{93 e^2}{2}+30 e^4 \right) 
+ (1-e^2)^{3/2} \left(\frac{261}{2}+\frac{1195 e^2}{4}-\frac{581 e^4}{8} \right) , \notag \\ 
\mathcal{U}_{7/2}^{S3} &= -2 (1-e^2)^2 (1 + e^2) + (1-e^2)^{3/2} \left(5+\frac{37 e^2}{2} \right)  
\end{align}
Note that any functions not explicitly referenced (both here and throughout this section), such as 
$\mathcal{U}_3^{S1}$ or $\mathcal{U}_{7/2}^{S2}$, are identically 0.

At 4PN order, the spin-independent portion becomes more complicated \cite{MunnEvan22a}; however, the spin
dependence remains simple:
\begin{align}
\mathcal{U}_4^{S2} &= (1-e^2)^2 ( 52+63 e^2-35 e^4) +(1-e^2)^{3/2} \left(-155-\frac{1095 e^2}{2}+21 e^4\right), 
\notag \\
\mathcal{U}_4^{S4} &= -3 (1-e^2)^{3/2} e^2
\end{align}
The 4.5PN functions can also be put into exact form, though with the first appearance of a transcendental
coefficient:
\begin{align}
\mathcal{U}_{9/2}^{S1} &= (1-e^2)^2 (-128-156 e^2-28 e^4+32 e^6)  \notag \\& \hspace{2em} 
+ (1-e^2)^{3/2} \left(\frac{5042}{9}
-\frac{241 \pi ^2}{96}+ \left(2699-\frac{405 \pi ^2}{32}\right)e^2 + \left(\frac{1625}{12}-\frac{569 \pi ^2}{256}\right)e^4
- \frac{1447 e^6}{8}\right) ,
\notag \\
\mathcal{U}_{9/2}^{S3} &= (1-e^2)^2 (-55+3 e^2+12 e^4) + (1-e^2)^{3/2} \left(105+\frac{867 e^2}{2}
+\frac{309 e^4}{4}\right) .
\end{align}
The 5PN functions likewise resemble their 4.5PN counterparts above, involving factors of $\pi^2$.
\begin{align}
\mathcal{U}_{5}^{S2} &= (1-e^2)^2 \left(\frac{1099}{2}-420 e^2+\frac{879 e^4}{2}-121 e^6\right) 
\notag \\& \hspace{2em} 
+ (1-e^2)^{3/2} \left( -\frac{7067}{6}+\frac{593 \pi ^2}{512}+ \left(-\frac{60485}{12}+\frac{3091 \pi ^2}{512}\right) e^2
+ \left(-\frac{4697}{3}+\frac{4403 \pi ^2}{4096}\right) e^4 +\frac{3689 e^6}{8}  \right), 
\notag \\
\mathcal{U}_{5}^{S4} &= (1-e^2)^2 (45-46 e^2+e^4) 
+ (1-e^2)^{3/2} \left(-53-\frac{319 e^2}{2}-\frac{127 e^4}{2} \right) .
\end{align}

The 5.5PN term $\mathcal{U}_{11/2}^{S1}$ marks the first appearance of additional transcendental numbers 
($\gamma_E, \log 2,$ etc.), as well as a $\log p$ term with spin dependence.  As might be expected from the 
Schwarzschild case, this is the first term for which we cannot determine a closed or exact function in eccentricity and 
must rely on the Taylor expansion through $e^{16}$.  However, we can factor this term into a simpler form 
reminiscent of $\mathcal{U}_{4}^{\rm Sch}$ \cite{MunnEvan22a}, in order to capture some of the transcendental 
dependence.  We present the infinite series portion of $\mathcal{U}_{11/2}^{S1}$ to $e^8$, saving the full results for
the online repositories \cite{BHPTK18, UNCGrav22}.  The remaining enhancement functions are closed in form.
\begin{align}
\mathcal{U}_{11/2}^{S1} &=  
\pi^2 \left(1-e^2\right)^{3/2} \left( -\frac{79573}{768}-\frac{50411 e^2}{128}-\frac{166217 e^4}{2048}
+\frac{4681 e^6}{6144} \right)  - 2 \left[\g_E +\log \left(\frac{8}{1+\sqrt{1-e^2}}\right) \right] \mathcal{U}_{11/2L}^{S1}
\notag \\& ~~
+ (1-e^2)^{3/2} \bigg[ \left(\frac{1163681}{450}-\frac{32 \log (2)}{15}\right)+\left(\frac{6101333}{300}
-\frac{45296 \log (2)}{15}  +2916 \log (3)\right) e^2   \notag \\& ~~
+\left(\frac{124628}{75} +\frac{729956 \log (2)}{15}-\frac{285039 \log (3)}{10}\right) e^4
\notag \\& ~ ~ +\left(-\frac{25544941}{7200}-\frac{17042578 \log (2)}{45}+\frac{1805733 \log (3)}{16}
 +\frac{12578125 \log (5)}{144}\right) e^6  \notag \\& ~~
 +\left(-\frac{61976903}{46080}+\frac{45460366 \log (2)}{27}+\frac{181311291 \log (3)}{2560}
 -\frac{10682734375 \log (5)}{13824}\right) e^8  + \cdots + \a_{16} e^{16}  + \mathcal{O}(e^{18}) \bigg]  , \notag \\
\mathcal{U}_{11/2}^{S3} &=  (1-e^2)^2 \left( -\frac{2399}{2}+\frac{5001 e^2}{4}-\frac{1843 e^4}{4}+130 e^6 \right) 
+ (1-e^2)^{3/2} \left( \frac{3289}{2}+\frac{11323 e^2}{2}+\frac{42641 e^4}{16}-\frac{4749 e^6}{16} \right) ,
\notag \\
\mathcal{U}_{11/2}^{S5} &= (1-e^2)^2 \left( -\frac{39}{2}+\frac{47 e^2}{2} \right) 
+ (1-e^2)^{3/2} \left( \frac{39}{2}+\frac{131 e^2}{4}+\frac{27 e^4}{2}  \right) , \notag \\
\mathcal{U}_{11/2L}^{S1} &= -(1-e^2)^{3/2} \left( \frac{1168}{15}+\frac{6584 e^2}{15}+\frac{1898 e^4}{5}
+\frac{491 e^6}{15}  \right) .
\end{align}

At 6PN order we find an analogous set of functions.  Again, we truncate the more complicated series at $e^{8}$,
leaving the full functions for the online repositories.  The remaining functions are likewise found to yield closed forms.
\begin{align}
\mathcal{U}_{6}^{S2} &=  
\pi^2 \left(1-e^2\right)^{3/2} \bigg[\frac{67439}{3072}+\frac{350857 e^2}{2048}+\frac{1564717 e^4}{8192}
+\frac{975293 e^6}{49152}+\left(-\frac{287}{32}-\frac{287 e^2}{64}\right) \left(1-e^2\right)^{3/2}
\bigg]   \notag \\& ~~
- 2 \left[\g_E +\log \left(\frac{8}{1+\sqrt{1-e^2}}\right) \right] \mathcal{U}_{6L}^{S2}
+ (1-e^2)^{3/2} \bigg[ \left(-\frac{741686}{225}+\frac{8 \log (2)}{5}\right)  \notag \\& ~~
+\left(-\frac{4131994}{75}+\frac{4508 \log (2)}{5}  -  \frac{4374 \log (3)}{5}\right) e^2  
+\left(-\frac{1275121}{75}-\frac{193109 \log (2)}{15}+\frac{296703 \log (3)}{40}\right) e^4   \notag \\& ~~
+\left(\frac{599647}{200}+\frac{7664441 \log (2)}{90}-\frac{3812427 \log (3)}{160}
-\frac{6015625 \log (5)}{288}\right) e^6    \notag \\&
+\left(\frac{2004023}{2560}-\frac{1426452 \log (2)}{5}-\frac{310330197 \log (3)}{10240}
+\frac{880390625 \log (5)}{6144}\right) e^8  + \cdots +  \b_{16} e^{16} + \mathcal{O}(e^{18})  \bigg]  , \notag \\
\mathcal{U}_{6}^{S4} &=  (1-e^2)^2 \left( 1444-1347 e^2-47 e^4-50 e^6 \right) 
+ (1-e^2)^{3/2} \left( -1600-4386 e^2-1908 e^4-\frac{339 e^6}{4} \right) ,
\notag \\
\mathcal{U}_{6}^{S6} &=  3 (1-e^2)^3 + (1 - e^2)^{3/2} \left( -3-\frac{9 e^2}{2} \right) , \notag \\
\mathcal{U}_{6L}^{S2} &= (1-e^2)^{3/2} \left( \frac{132}{5} + \frac{718 e^2}{5} + \frac{293 e^4}{2} + \frac{883 e^6}{60} \right) .
\end{align}

From 6.5PN through 8PN, our expansion is limited to order $e^{10}$ in eccentricity, which greatly reduces our ability
to determine closed-form expressions from the series.  From here, we primarily present just the first few coefficients
in each function.  Again, the full results are posted at the repositories \cite{BHPTK18, UNCGrav22}.  
The 6.5PN term marks the first appearance of the polygamma function.  We set 
$\psi^{(n, k)} = \psi^{(n)} (\frac{i k a}{\ka}) +  \psi^{(n)} (-\frac{i k a}{\ka})$ for polygamma function $\psi^{(n)}(x)$
and find
\begin{align}
\mathcal{U}_{13/2}^{S1} &= -\frac{1}{3} (1-e^2)^{3/2} \left( \log \ka - \g_E + \frac{1}{2} \psi^{(0, 2)} \right) 
\mathcal{U}_{13/2L}^{S3} - 2 \left[\g_E +\log \left(\frac{8}{1+\sqrt{1-e^2}}\right) \right] \mathcal{U}_{13/2L}^{S1}
\notag \\&
+ (1-e^2)^{3/2} \bigg[\left(\frac{137967017}{4410}-\frac{2648651 \pi ^2}{1024} - \frac{1528 \log (2)}{5}
+\frac{1944 \log (3)}{7}\right)   \notag \\&
+\left(\frac{3227202952}{11025}-\frac{27517957 \pi ^2}{1536}
+\frac{88816 \log (2)}{15}+\frac{5346 \log (3)}{7}\right) e^2  
+\bigg(\frac{4589364091}{29400}  \notag \\&
-\frac{146807929 \pi ^2}{12288}-\frac{6151762 \log (2)}{35}-\frac{5423031 \log (3)}{140}
+\frac{5859375 \log (5)}{56}\bigg) e^4 + \cdots + \g_{10} e^{10} + \mathcal{O}(e^{12}) \bigg]  ,  \notag \\
\mathcal{U}_{13/2}^{S3} &= - (1-e^2)^{3/2} \left( \log 2 \ka + \g_E + \frac{1}{2} \psi^{(0, 2)} \right) 
\mathcal{U}_{13/2L}^{S3} 
+ (1-e^2)^{3/2} \bigg[\left(\frac{696161}{225}-\frac{115 \pi ^2}{96}\right)  \notag \\&
+\left(\frac{1731571}{18}-\frac{38825 \pi ^2}{768}\right) e^2  
+\left(\frac{776917}{24}-\frac{17911 \pi ^2}{512}\right) e^4
+\left(-\frac{119507}{48}-\frac{3483 \pi ^2}{2048}\right) e^6   \notag \\&
+\left(-\frac{162431}{256}+\frac{861 \pi ^2}{2048}\right) e^8
+\left(-\frac{47309}{1600}+\frac{369 \pi ^2}{2048}\right) e^{10}+\mathcal{O}(e^{12}) \bigg]  ,  \notag \\
\mathcal{U}_{13/2}^{S5} &= (1-e^2)^{3/2} \left( \frac{102}{5}+3694 e^2+\frac{2503 e^4}{4}
-\frac{333 e^6}{8}-\frac{6157 e^8}{128}-\frac{345 e^{10}}{16} + \mathcal{O}(e^{12}) \right) , \notag \\
\mathcal{U}_{13/2L}^{S1} &= - (1-e^2)^{3/2} \left( \frac{18268}{105}+\frac{31008 e^2}{35}+\frac{72857 e^4}{105}
+\frac{2118 e^6}{7}+\frac{62603 e^8}{1120} \right) , \notag \\
\mathcal{U}_{13/2L}^{S3} &= -(1-e^2)^{3/2} \left( \frac{96}{5}+144 e^2+108 e^4+6 e^6 \right) .
\end{align}

The 7PN functions are similar in structure to their 6PN counterparts, though we do note the first appearance of an 
odd power of $a$ at integer order.  Additionally, we no longer have enough coefficients in the function $S2$ to 
identify any eccentric structure, so we simply present a few of those coefficients unmodified.  We find
\begin{align}
\mathcal{U}_{7}^{S1} &= \pi (1-e^2)^{3/2} \left( \frac{343088}{1575}+\frac{394122 e^2}{175}
+\frac{5969582 e^4}{1575}+\frac{661759039 e^6}{453600}
+\frac{152835577 e^8}{2419200}+\frac{246822697 e^{10}}{290304000} + \mathcal{O}(e^{12}) \right)  \notag \\
\mathcal{U}_{7}^{S2} &= (1-e^2)^{3/2} \bigg[ \left(\frac{15442453}{3150}-\frac{74024 \g_E }{105}
-\frac{193510709 \pi ^2}{98304}-\frac{27016 \log (2)}{21}-\frac{729 \log (3)}{7}\right)   \notag \\& ~~
+\left(-\frac{633681119}{2100} -\frac{685264 \g_E }{105}-\frac{736121519 \pi^2}{49152} 
+\frac{322384 \log(2)}{35}  -\frac{785376 \log (3)}{35}\right) e^2   \notag \\& ~~
+\left(-\frac{140377429}{1680}-\frac{1019254 \g_E }{105}
-\frac{4127285485 \pi^2}{393216}-\frac{46263526 \log (2)}{105}+\frac{327468987 \log (3)}{1120}
-\frac{21484375 \log (5)}{672}\right) e^4   \notag \\& ~~
+ \cdots + \d_{10} e^{10} + \mathcal{O}(e^{12})\bigg]  \notag \\
\mathcal{U}_{7}^{S4} &= (1-e^2)^{3/2} \bigg[ \left(-\frac{4900}{3}-\frac{69 \pi ^2}{256}\right)+\left(-115532
+\frac{585 \pi ^2}{64}\right) e^2+\left(-\frac{313281}{8}+\frac{10521 \pi ^2}{2048}\right) e^4  \notag \\& ~~
+\left(\frac{129643}{48}-\frac{267 \pi ^2}{2048}\right) e^6+\left(\frac{119151}{128}-\frac{861 \pi^2}{8192}\right) e^8
+\left(\frac{108689}{256}-\frac{369 \pi ^2}{8192}\right) e^{10}+\mathcal{O}(e^{12}) \bigg] ,  \notag \\
\mathcal{U}_{7}^{S6} &= (1-e^2)^{3/2} \bigg( -1374 e^2-\frac{2145 e^4}{16}+\frac{1671 e^6}{32}
+\frac{5055 e^8}{256}+\frac{4401 e^{10}}{512}+\mathcal{O}(e^{12})  \bigg)  ,  \notag \\
\mathcal{U}_{7L}^{S2} &= (1-e^2)^{3/2} \bigg(\frac{37012}{105}+\frac{342632 e^2}{105}+\frac{509627 e^4}{105}+
\frac{206029 e^6}{105}+\frac{27501 e^8}{224}-\frac{2303 e^{10}}{240}  + \mathcal{O}(e^{12}) \bigg)
\end{align}

The 7.5PN functions $S1$ and $S3$ show interesting behavior, in that they present certain transcendental numbers
attached to (apparently) terminating series that do not bear any obvious relationship to the corresponding 
logarithmic functions.
\begin{align}
\mathcal{U}_{15/2}^{S1} &= (1-e^2)^{3/2} \bigg[ \left(\frac{224}{5}+\frac{2448 e^2}{5}+708 e^4+182 e^6
+\frac{9 e^8}{2} \right) \log \ka + \left( \frac{16}{5}+\frac{224 e^2}{5}+84 e^4+28 e^6+\frac{7 e^8}{8} \right)
\psi^{(0,1)}    \notag \\&
+ \left( \frac{96}{5}+200 e^2+270 e^4+63 e^6+\frac{11 e^8}{8} \right) \psi^{(0,2)} + \bigg(\frac{8776579}{32768}
+\frac{653303597 e^2}{262144}+\frac{2529471329 e^4}{1048576}+\frac{174272943 e^6}{2097152}    \notag \\&
-\frac{317401461 e^8}{16777216} \bigg) \pi^4 \bigg]     
+(1-e^2)^{3/2}  \bigg[ \bigg(\frac{630173174963}{3572100}-\frac{2026048 \g_E }{2835}
-\frac{15709506835 \pi ^2}{884736}  -  \frac{259648 \log (2)}{2835}   \notag \\&  
-\frac{5832 \log (3)}{7} \bigg) 
+\bigg(\frac{6701813671607}{2381400}-\frac{889124 \g_E }{81}-\frac{97156196179 \pi ^2}{442368}  
-\frac{429451756 \log (2)}{2835}+\frac{5378319 \log(3)}{140}     \notag \\&
+\frac{9765625 \log (5) }{324} \bigg) e^2  +  \cdots + \e_{10} e^{10} + \mathcal{O}(e^{12}) \bigg]  , \notag \\
\mathcal{U}_{15/2}^{S2} &= \pi (1-e^2)^{3/2} \bigg( -\frac{5564}{105}-\frac{81213 e^2}{175} - \frac{1577929 e^4}{1800}
 -\frac{5597705 e^6}{12096} - \frac{9631177 e^8}{387072} - \frac{1386265357 e^{10}}{2903040000}
 +\mathcal{O}(e^{12}) \bigg)  ,  \notag \\
 \mathcal{U}_{15/2}^{S3} &= (1-e^2)^{3/2} \bigg[ \left(\frac{552}{5}+\frac{5664 e^2}{5}+1494 e^4+336 e^6
+\frac{111 e^8}{16} \right) \log \ka - \left( \frac{12}{5}+\frac{168 e^2}{5}+ 63 e^4 + 21 e^6+ \frac{21 e^8}{32} \right)
\psi^{(0,1)}    \notag \\& ~~
+ \left(\frac{288}{5}+600 e^2+810 e^4+189 e^6+\frac{33 e^8}{8} \right) \psi^{(0,2)} \bigg] 
+(1-e^2)^{3/2}  \bigg[ \bigg(\frac{5268133}{225}+\frac{6488 \g_E }{15}-\frac{3298493 \pi ^2}{6144}  \notag \\& ~~
+\frac{11096 \log (2)}{15}\bigg)  +   \left(\frac{246222767}{225}+\frac{68968 \g_E }{15}
-\frac{4438835 \pi ^2}{768}-6120 \log (2)+\frac{64152 \log (3)}{5}\right) e^2  \notag \\& ~~
+  \cdots + \xi_{10} e^{10} + \mathcal{O}(e^{12}) \bigg] ,  \notag \\
 \mathcal{U}_{15/2}^{S5} &= (1-e^2)^{3/2} \left(\frac{2251}{5}+\frac{490316 e^2}{5}+\frac{64077 e^4}{2}
 -\frac{53317 e^6}{16}-\frac{605545 e^8}{512}-\frac{325199 e^{10}}{512}  + \mathcal{O}(e^{12}) \right), 
 \notag \\ 
 \mathcal{U}_{15/2}^{S7} &= (1-e^2)^{3/2} \left( 294 e^2+\frac{7 e^4}{2}-\frac{241 e^6}{16}-\frac{849 e^8}{256}
  -\frac{11 e^{10}}{8}  + \mathcal{O}(e^{12}) \right), 
 \notag \\ 
\mathcal{U}_{15/2L}^{S1} &= (1-e^2)^{3/2} \left( \frac{189904}{567}+\frac{2123666 e^2}{405}
+\frac{50236532 e^4}{2835}+\frac{27140887 e^6}{1890}+\frac{329151 e^8}{280}
-\frac{5671807  e^{10}}{20160} +\mathcal{O}(e^{12}) \right)  , \notag \\ 
\mathcal{U}_{15/2L}^{S3} &= - (1-e^2)^{3/2} \left( \frac{4072}{15}+\frac{8596 e^2}{3}+\frac{21511 e^4}{5}
+\frac{16149 e^6}{10}+\frac{1213 e^8}{15}-\frac{329 e^{10}}{40} +\mathcal{O}(e^{12}) \right)  .
\end{align}

Finally, the 8PN functions introduce several elements of additional new structure.  One is in the form of a
spin-dependent term $\mathcal{U}_{8}^{S0}$ with no leading factor of $a$.  The attached eccentricity series for both
$\mathcal{U}_{8}^{S0}$ and $\mathcal{U}_{8}^{S3}$ appear to terminate at $e^8$, so they presented in their entirety,
while the (apparently infinite) series $\mathcal{U}_{8}^{S1}, \mathcal{U}_{8}^{S2},$ and $ \mathcal{U}_{8}^{S4}$ are 
truncated for brevity.  Additionally, we find the appearance of a second polygamma combination, commonly denoted
$\bar{\psi}^{(n,k)} = -i (\psi^{(n)}( \frac{i k a}{\ka}) - \psi^{(n)}(-\frac{i k a}{\ka}))$, as well of $\ka$ in the denominators 
of certain coefficients.  The full results begin
\begin{align}
\mathcal{U}_{8}^{S0} &= (1-e^2)^{3/2} \bigg[ \left(\frac{1712}{525}-\frac{64 \g_E }{5}+\frac{1712 \ka}{525}
-\frac{32 \psi^{(0,2)}}{5} -\frac{64 \log (\ka)}{5}\right)+\bigg(\frac{26536}{525}-\frac{968 \g_E }{5}  
+\frac{25252 \ka}{525} -\frac{484 \psi^{(0,2)}}{5}   \notag \\& ~~
-\frac{992 \log (\ka)}{5}\bigg) e^2+\left(\frac{3638}{35}-390 \g_E 
+\frac{3317 \ka}{35} -195  \psi^{(0,2)}-408 \log (\ka)\right) e^4+\bigg(\frac{3959}{105}-139 \g_E   \notag \\& ~~
+\frac{1391 \ka}{42}
-\frac{139 \psi^{(0,2)}}{2}-148 \log (\ka)\bigg) e^6+\left(\frac{107}{84}-\frac{37 \g_E }{8}+\frac{1819 \ka}{1680}
-\frac{37 \psi^{(0,2)}}{16}-5 \log (\ka)\right) e^8  \bigg] ,  \notag \\
\mathcal{U}_{8}^{S1} &= (1-e^2)^{3/2} \bigg[ \left(\frac{2389021 \pi }{9450}-\frac{3424 \bar{\psi}^{(0,2)}}{525}
-\frac{192 \ka \bar{\psi}^{(1,2)}}{5}+\frac{256 \bar{\psi}^{(1,2)}}{5 \ka}\right)  \notag \\& ~~
+\left(-\frac{108802823 \pi }{132300}-\frac{51788 \bar{\psi}^{(0,2)}}{525}-\frac{2904 \ka \bar{\psi}^{(1,2)}}{5}
+\frac{3872 \bar{\psi}^{(1,2)}}{5 \ka}\right) e^2 + \cdots + \chi_{10} e^{10} + \mathcal{O}(e^{12}) \bigg] , \notag \\
\mathcal{U}_{8}^{S2} &= (1-e^2)^{3/2} \bigg[ \bigg(\frac{75054389639}{198450}-\frac{1753576 \g_E }{567}
+\frac{1712 \ka}{175}
-\frac{314206240595 \pi^2}{7077888}-\frac{417436343 \pi ^4}{16777216}-\frac{64 \psi^{(0,1)}}{15}   \notag \\& ~~
-\frac{432 \psi^{(0,2)}}{5}-\frac{14672264 \log (2)}{2835}-972 \log (3)-\frac{544 \log (\ka)}{3}\bigg)
+\bigg(\frac{29407415438}{14175}-\frac{3574292 \g_E }{81}+\frac{26322 \ka}{175}   \notag \\& ~~
-\frac{8172744049193 \pi ^2}{14155776}-\frac{1047412729 \pi ^4}{4194304}-\frac{896 \psi^{(0,1)}}{15}
-1276 \psi^{(0,2)}+\frac{134696372 \log (2)}{2835}-\frac{36823977 \log (3)}{280}    \notag \\& ~~
-\frac{48828125 \log (5)}{4536} -\frac{39928 \log (\ka)}{15}\bigg) e^2 + \cdots + h_{10} e^{10} 
+ \mathcal{O}(e^{12}) \bigg]  ,  \notag \\
\mathcal{U}_{8}^{S3} &= (1-e^2)^{3/2} \bigg[ -\frac{3424 \bar{\psi}^{(0,2)}}{175} 
-\frac{51788 \bar{\psi}^{(0,2)} e^2}{175}-\frac{4173 \bar{\psi}^{(0,2)} e^4}{7}
-\frac{14873 \bar{\psi}^{(0,2)} e^6}{70}-\frac{3959 \bar{\psi}^{(0,2)} e^8}{560} \bigg]  ,   \notag \\
\mathcal{U}_{8}^{S4}  &= (1-e^2)^{3/2} \bigg[ \left(-\frac{2910653}{225}-\frac{584 \g_E }{5}
+\frac{269059 \pi ^2}{98304}+\frac{16 \psi^{(0,1)}}{5}-48 \psi^{(0,2)}-\frac{712 \log(2)}{5}-\frac{448 \log (\ka)}{5}\right)
\notag \\ &~~
+\bigg(-\frac{81173087}{50}-1624 \g_E -\frac{214 \ka}{175} +\frac{9737469 \pi^2}{16384}
+\frac{224 \psi^{(0,1)}}{5}-\frac{3168 \psi^{(0,2)}}{5}-\frac{464 \log (2)}{5}-\frac{8748 \log (3)}{5}  \notag \\&
-\frac{5912 \log (\ka)}{5}\bigg) e^2 + \cdots + q_{10} e^{10} + \mathcal{O}(e^{12}) \bigg]  ,  \notag \\
\mathcal{U}_{8}^{S6}  &= (1-e^2)^{3/2} \bigg( -\frac{249}{5}-\frac{293474 e^2}{5}-17592 e^4+\frac{12229 e^6}{4}
+\frac{268889 e^8}{256}+\frac{121663  e^{10}}{256}  +  \mathcal{O}(e^{12})  \bigg) ,  \notag \\ 
\mathcal{U}_{8}^{S8}  &= (1-e^2)^{3/2} \bigg( -27 e^2+\frac{27 e^4}{8}+\frac{17 e^6}{16}+\frac{9 e^8}{32}
+\frac{33 e^{10}}{256} +  \mathcal{O}(e^{12})  \bigg)  ,  \notag \\
\mathcal{U}_{8L}^{S2}  &= (1-e^2)^{3/2} \bigg( \frac{928196}{567}+\frac{9472814 e^2}{405}
+\frac{192633277 e^4}{5670}+\frac{86757829 e^6}{11340}+\frac{22396051 e^8}{10080}
+\frac{3088957 e^{10}}{10080}+  \mathcal{O}(e^{12})  \bigg) ,  \notag \\
\mathcal{U}_{8L}^{S4}  &= (1-e^2)^{3/2} \bigg(\frac{516}{5} + \frac{7028 e^2}{5}+\frac{40949 e^4}{15}
+\frac{16568 e^6}{15}+\frac{21739 e^8}{480}-\frac{329 e^{10}}{160}+  \mathcal{O}(e^{12})  \bigg)  .
\end{align}

\end{widetext}
\subsection{Discussion}
\label{sec:ResultDiss}

When PN expansions are made in the extreme-mass-ratio limit using the Teukolsky-MST formalism described in this 
paper, the result is a double Taylor series about $1/p = 0$ and $e = 0$, taken to finite orders in both parameters.  
Nevertheless, significant prior
work at the intersection of BHPT and PN theory in the Schwarzschild limit has revealed that the derived terms should
have significant structure in their dependence on eccentricity.  Direct derivations from the full PN theory, for instance,
have found closed-form expressions (or fully understood infinite series) for the first three orders in the energy flux
and the redshift invariant \cite{PeteMath63, ArunETC08a, ArunETC08b, AkcaETC15}.  Likewise, work has been
done to characterize the behavior of eccentricity enhancement functions as $e$ approaches 1, again in the 
Schwarzschild limit \cite{ForsEvanHopp16, LoutYune17, MunnETC20}.  Our knowledge of the underlying PN 
structure can then be used to refactor the initial (Taylor series) results into the corresponding exact functions of 
eccentricity that would produce those series, greatly enhancing our access to the high-$e$ regime.  

In the Schwarschild case, this effort proved highly effective, as BHPT-PN expansions were found yield a great 
many closed forms in the fluxes \cite{MunnEvan19a, MunnEvan20a, Munn20, MunnEvanFors23}, redshift invariant
\cite{MunnEvan22a}, and spin-precession invariant \cite{MunnEvan22b}.  With the success of those methods in 
hand, we were motivated to push the eccentric expansion in the Kerr regime, to see if similar structure would
appear in the spin-dependence of the PN expansion.  We first confirmed some of the same structure in the fluxes
at infinity \cite{MunnETC23}, and with this work we now see that many of the patterns repeat in the redshift
invariant, though not without significant added complexity and a few unexpected irregularities.  We review the
results here.

The first eight PN functions, $\mathcal{U}_{3/2},$ $\mathcal{U}_{2},$ $\mathcal{U}_{5/2},$ $\mathcal{U}_{3},$ 
$\mathcal{U}_{7/2},$ $\mathcal{U}_{4},$ $\mathcal{U}_{9/2},$ and $\mathcal{U}_{5}$, are all found to yield straightforward 
closed forms.  When separated by power of $a$, we find that each takes the form of a pair of polynomials in $e$, each 
attached to a factor of $(1-e^2)$ to some positive power.  (Recall that when $y=(M \O_\vp)^{2/3}$ is used as the PN 
expansion parameter instead of $1/p$, the factors of $(1-e^2)$ carry negative powers, commonly called eccentricity 
singular factors \cite{ForsEvanHopp16, MunnEvan19a}.)  These functions are all vaguely reminiscent of the 2PN energy
flux \cite{ArunETC08a}, which was the first derived PN term to display a similar combination of polynomials attached 
to differing eccentricity singular factors.  

Using the Schwarzschild limit as a model, it could perhaps be predicted that these functions would be compact, as 
the first term there with a nontrivial structure was $\mathcal{U}_{4}^{\rm Sch}$.  Thus, we might expect that the first 
nontrivial term in the spinning case would be four orders after the first, at 5.5PN order.  Nevertheless, it is 
remarkable that the spin dependence (at least in the equatorial limit) can be known exactly up to that point.  
Additionally, the only term in the Schwarzschild limit not fully understood to this level is $\mathcal{U}_{5}^{\rm Sch}$, 
as it has a dependence on the 1PN multipole moments that has not yet been elucidated (see \cite{MunnEvan22a} for 
more details).  Once that piece is determined, the entire 5PN redshift invariant will be understood for eccentric 
equatorial EMRIs on a Kerr background.

The 5.5PN function $\mathcal{U}_{11/2}^{S1}$ carries the first major increase in coefficient complexity, presenting a form 
that resembles $\mathcal{U}_{4}^{\rm Sch}$.  Given the similarity, and based on prior 
work in the Schwarzschild limit \cite{MunnEvan20a, MunnEvan22a}, we would expect the log coefficients in this term 
to stem from a $\chi(e)$-like enhancement function (see \cite{ArunETC08a,ArunETC08b,ArunETC09a}), meaning 
they cannot be condensed into closed form.  Nevertheless, if this is the case, such a function would likely yield to a 
derivation in terms of multipole moments, which could be expanded to arbitrary order in eccentricity.  Indeed, this was 
precisely found to be the case for $\mathcal{U}_{4}^{\rm Sch}$ \cite{MunnEvan22a}.  Despite a structure identical to that of 
$\mathcal{U}_{11/2}^{S1}$, knowledge of the multipolar composition of $\mathcal{U}_{4}^{\rm Sch}$ allows us to expand the 
function to arbitary order in eccentricity.

Further bolstering support for this prospect is the fact that the 5.5PN log term $\mathcal{U}_{11/2L}^{S1}$ is
exactly proportional to the leading spin-dependence of the energy flux $\mathcal{L}_{3/2}^{S1}$ \cite{MunnETC23}.  
Curious connections between the energy flux series and redshift invariant series also appeared in the Schwarzschild 
limit \cite{MunnEvan22a}.  In fact, we found that the entire leading logarithm series of the energy flux (that is, the first
appearance of each new power of $\log p$, including power 0 --- see \cite{MunnEvan19a}) recurred in the redshift 
invariant, but shifted four PN orders up.  Indeed, the Peters-Mathews energy flux term was found to be proportional to 
$\mathcal{U}_{4L}^{\rm Sch}$.  In this case we could not find an infinite sequence of PN terms with proportionality 
between the two, but the three terms $\mathcal{U}_{11/2L}^{S1}$, $\mathcal{U}_{6L}^{S2}$, and 
$\mathcal{U}_{15/2}^{S2}$ showed correspondence with $\mathcal{L}_{3/2}^{S1}$, $\mathcal{L}_{2}^{S2}$, and 
$\mathcal{L}_{7/2}^{S2}$, respectively.  Whether there is a deeper pattern there remains an open question and will
be left to future investigation.

The next major increase in complexity occurs at 6.5PN order, with the functions $\mathcal{U}_{13/2}^{S1}$ and
$\mathcal{U}_{13/2}^{S3}$.  We have used the limited number of coefficients to extract what seems to be the 
complete dependence of the polygamma, $\log \ka$, and $\g_E$ terms on the logarithmic functions 
$\mathcal{U}_{13/2L}^{S1}$ and $\mathcal{U}_{13/2L}^{S3}$.  The fact that both different logarithmic eccentricity functions 
are present in the non-logarithmic terms $\mathcal{U}_{13/2}^{S1}$ is curious, and we conjecture that it 
represents the effect of a $\ka^2$ term on a combination of coefficients at an earlier stage of the derivation.  
The term $\mathcal{U}_{13/2}^{S3}$ also exposes an additional irregularity, as it is the third PN order with an $a^3$
dependence, meaning we would naively (following the pattern in the Schwarzschild, $S1$, and $S2$ functions) 
expect a series with no transcendentals other than $\pi^2$.  However, we instead find further appearances of 
polygamma, $\log \ka$, and $\g_E$, again possibly resulting from $\ka^2$ terms in the MST solutions.

At 7PN we see the first occurrence of an $S1$ function at an integer order.  This function is attached to a factor of 
$\pi$, which is reminiscent of low-order tail terms in the energy flux.  This makes sense, as the half-integer tail for the 
redshift invariant starts at 5.5PN in the Schwarzschild limit, and $\mathcal{U}_{11/2}^{\rm Sch}$ takes a very similar 
form to $\mathcal{U}_{7}^{S1}$.  It is unlikely that this series terminates at finite order, but if its multipolar content resembles
that of tail contributions to other observables like the fluxes, there may a route to an arbitrary-order expansion.  The higher 
spin terms follow the same general structure as their counterparts at 6PN, though the
lower range of the eccentricity expansion does not provide us with enough information to determine closed or
exact functions.  It is perhaps noteworthy that the 7PN functions are markedly simpler than their 6.5PN counterparts, 
showing no incidence of polygamma or $\log \ka$.  Additionally, the logarithmic term only contains one 
$a$-function in $\mathcal{U}_{7L}^{S2}$, while $\mathcal{U}_{13/2L}$ produced both $S1$ and $S3$.  

There are a few interesting functions at 7.5PN order as well.  In particular, in the functions $S1$ and $S3$, we 
find the transcendental terms $\log \ka$, $\psi^{(0,1)}$, and $\psi^{(0,2)}$, all attached to eccentricity series that
appear to terminate at $e^8$.  Moreover, the polynomials are linearly independent and bear no apparent relationship
to the logarithmic functions $\mathcal{U}_{15/2L}^{S1}$ and $\mathcal{U}_{15/2L}^{S3}$.  As mentioned above, it 
is also noteworthy that the function $\mathcal{U}_{15/2}^{S2}$, the first appearance of an even power of $a$ at
half-integer order, shows the last connection to the energy flux expansion.  Indeed the infinite eccentricity series is exactly
proportional to the 3.5PN energy flux term $\mathcal{L}_{7/2}^{S2}$, though the more fundamental reason behind this 
particular connection remains unknown.

The final term at 8PN introduces another combinatorial increase in complexity.  8PN is the first order with $a$ 
dependence in the term $S0$.  In fact, this function appears to have terms that are independent of $a$ and therefore
belong in the Schwarzschild limit.  However, it turns out that when the limit is taken $a \rightarrow 0$, these terms 
exactly cancel with polygamma terms that remain nonzero in $\mathcal{U}_{8}^{S1}$.  Additionally, the term 
$\mathcal{U}_{8}^{S0}$ appears to truncate at $e^8$.  The functions $S1$ through $S4$ display similar behavior (though 
with varying degrees of complexity), while the 
remaining functions through $\mathcal{U}_{8}^{S6}$, $\mathcal{U}_{8}^{S8}$, $\mathcal{U}_{8L}^{S2}$, and
$\mathcal{U}_{8L}^{S4}$ are rational series.  Broadly speaking, it appears the maximum power of $a$ present in a 
given PN term tends to follow separate patterns for half-integer and integer orders.  The half-integer orders follow the 
trend $\{1,1,3,3,5,5\cdots\}$, while the integer orders follow the trend $\{2,2,4,4,6,6, \cdots\}$.  At each order, the 
higher powers of $a$ tend to be rational, providing ample opportunity for the derivation of closed-form expressions.

\subsection{Comparison to numerical data}

\begin{figure*}
\hspace{-1.5em}\includegraphics[scale=.71]{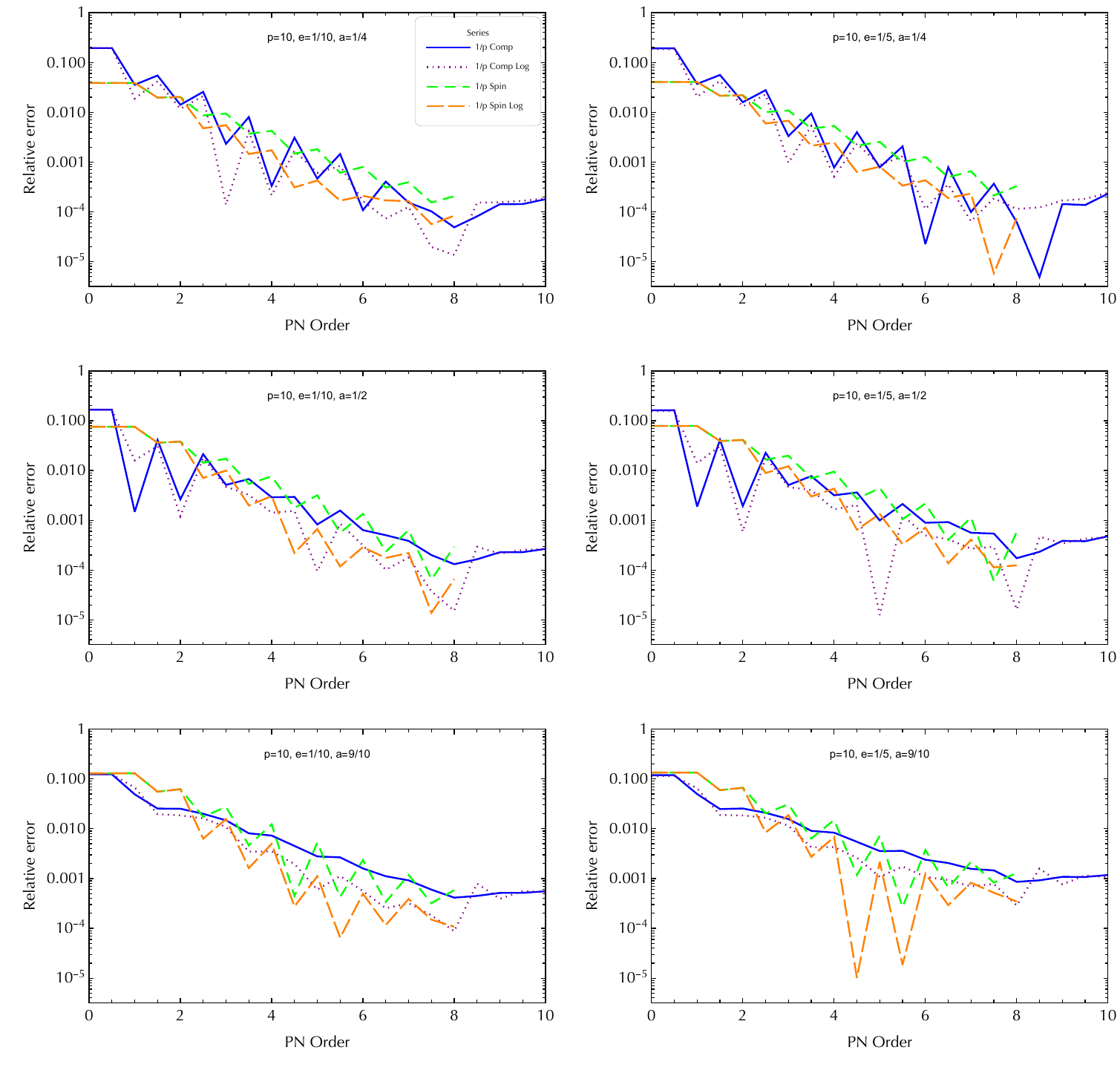}
\caption{Accuracy of the redshift invariant PN expansion and its resummations for several individual orbits.  The
numerical values of our redshift expansion are plotted against numerical calculations for several orbits with $p=10$. 
Within each plot comparisons are made for a composite (``Comp") expansion against the full numerical redshift, as well
as for the spin-dependent portion of the expansion against the redshift's spin-dependent residual, both with and without the 
use of the logarithmic summation.   Numerical data supplied by Zachary Nasipak.  Lines in the plots level off when the 
expansion is accurate to within numerical error bounds.
\label{fig:p10comps}}
\end{figure*}

\begin{figure*} 
\hspace{-1.5em}\includegraphics[scale=.71]{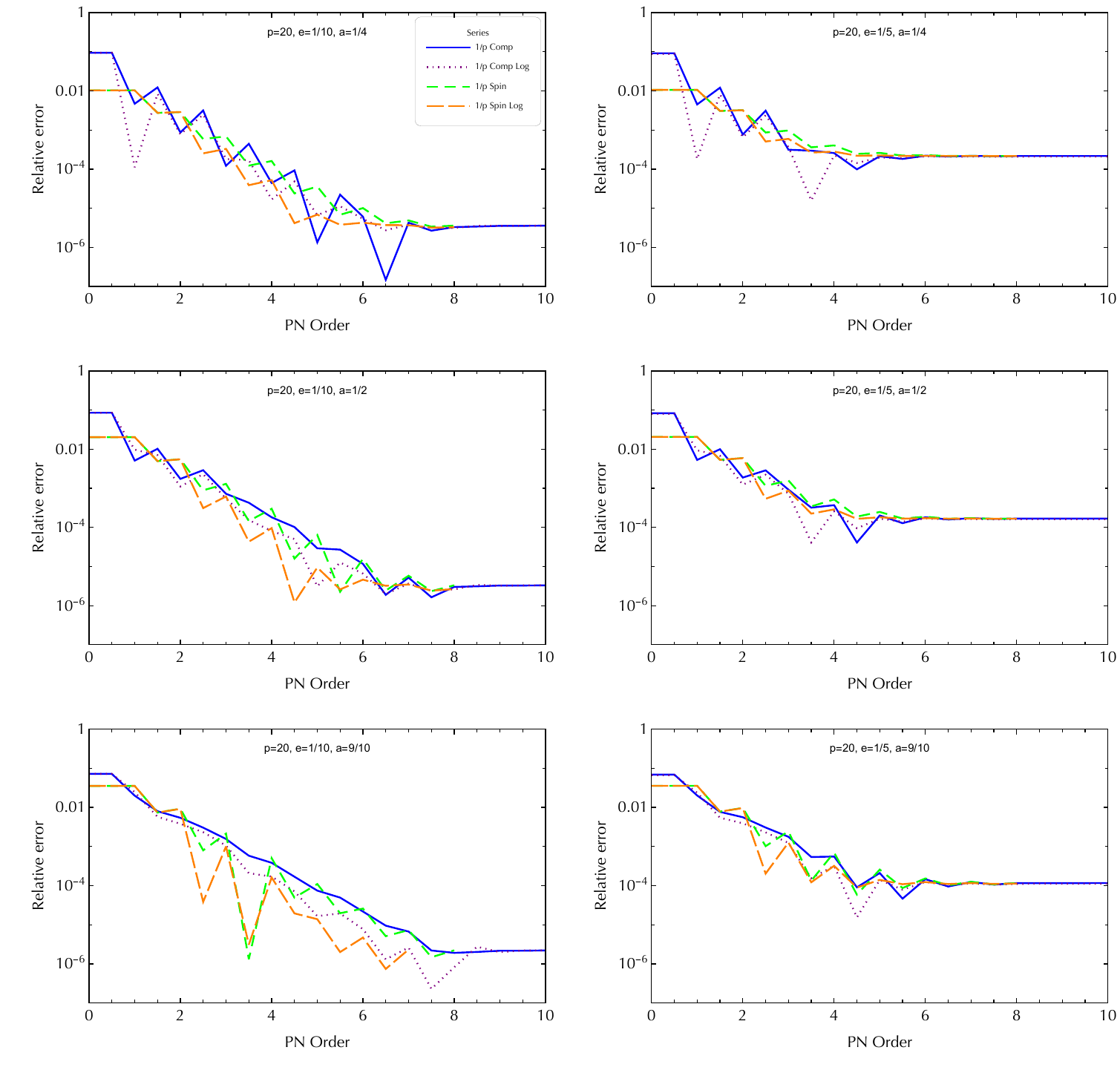}
\caption{Accuracy of the redshift invariant PN expansion and its resummations for $p=20$.  Numerical data supplied by 
Zachary Nasipak.  Lines in the plots level off when the expansion is accurate to within numerical error bounds (in particular,
data points for $p=20, e=1/5$ were computed to lower accuracy, leading to earlier level off).
\label{fig:p20comps}}
\end{figure*}

We can assess the accuracy of these expansions by comparing them to the numerical calculations of the redshift invariant for 
specific values of $p,$ $e,$ and $a$.  We evaluate orbits with $p=\{10, 20\}$, $e=\{1/10, 1/5\}$, and $a=\{1/4, 1/2, 9/10\}$ in 
an attempt to survey a range of parameters.  The spin-independent portion of the expansion
was more completely determined in \cite{MunnEvan22a}, so we supplement the current equatorial Kerr series with additional
coefficients from \cite{MunnEvan22a} as needed.  In order to better understand the behavior of the series, we make 
comparisons in two ways: 1) We construct a composite series, in which the spin-independent portion is supplied by 
the results in \cite{MunnEvan22a} to 10PN, and the spin-dependent portion from the present calculation is added through 
8PN.  2) We use only the spin-dependent portion of the PN series, and we compare this to the residual numerical calculation 
found by subtracting the Schwarzschild redshift off the full Kerr value.   Note that in this case, the 
fractional error is still computed with respect to the full (Kerr) redshift value, not the difference.  Finally, we try the logarithmic
resummation of each method to check its effects on convergence \cite{IsoyETC13, JohnMcDa14}.  These results are 
presented for $p=10$ in Fig.~\ref{fig:p10comps} and for $p=20$ in Fig.~\ref{fig:p20comps}.

From the plots, we can see that the convergence follows a few trends across values of $p,$ $e,$ $a,$ and PN order.  The PN 
regime corresponds to larger $p$ by definition, so the reduction in error from $p=10$ to $p=20$ matches expectations.  The
plots for the orbits with $p=20, e=1/5$ reach the numerical error threshold around 6PN order, but we can still observe a 
steeper trend in these graphs than in their $p=10$ counterparts.  Similarly, experience in the Schwarzschild limit 
\cite{MunnEvan22a, Munn20} has revealed that the PN convergence tends to worsen with increasing $e$, which is generally 
reflected from the left to the right columns of the plots (though this fact is somewhat obscured by occasional irregular jumps).
What is perhaps most interesting is the apparent loss of convergence with increasing $a$.  Higher values of $a$ permit stable 
orbits with smaller values of $p$, so we might well expect higher $a$ to improve convergence against the same value of $p$.
The fact that the opposite happens (at least in these sample orbits) could indicate weakened utility for the PN expansion in 
the high spin regime.

Beyond these trends, we can note that each orbit with sufficient numerical accuracy shows monotonic improvement until 
around the 8PN level, at which point the spin-dependence is lost.  This fact implies that the Schwarzschild portion of the
expansion, which remains from 8PN-10PN, is a poor substitute for the composite expansion.  On the other hand, the 
steady improvement prior to that point implies that accuracy could continue to improve through the use of higher-order
expansions.  The present work to 8PN and $e^{10}$ reached the limit of our supercomputing resources, with the bottleneck
step requiring many parallelized jobs each lasting 8-10 days on the UNC supercomputing cluster Longleaf.  Nevertheless, it 
would certainly be possible to extend the PN order at the expense of eccentricity, permitting an expansion to, say, 10PN and 
$e^4$.  As usual, we should recall that the contributions to the orbital phase evolution by conservative terms are suppressed 
by the mass ratio relative to the flux \cite{HindFlan08}, implying that even a slow-to-converge PN expansion 
of the conservative part of the self-force may be useful in close orbits.

Finally, we note that the comparison of the spin-dependence alone against the residual difference between the Kerr
redshift and the Schwarzschild redshift did not improve the convergence compared to the simple composite expansion.  
However, the residual expansion is very simple in form, so there may be a computational advantage to approaching the 
problem in this manner.

\section{Conclusions}
\label{sec:ConsConc}

This work has analytically derived the PN expansion of the generalized redshift invariant for eccentric, equatorial EMRIs with 
a Kerr primary to high order.  The series is computed to 8PN and $e^{10}$ in eccentricity, with the PN terms through 6PN 
found to $e^{16}$.  Most importantly, each term in the expansion retained exact dependence on the spin parameter $a$, 
greatly advancing past work in the small-$a$ limit \cite{BiniGera19a}.  The depth of the eccentricity expansion allows us to 
resum several eccentricity terms into closed-form expressions.  Explicitly, exact expressions were found for the
eccentricity-dependence (and spin-dependence) of the full terms $\mathcal{U}_{3/2},$ $\mathcal{U}_{2},$ 
$\mathcal{U}_{5/2},$ $\mathcal{U}_{3},$ $\mathcal{U}_{7/2},$ $\mathcal{U}_{4},$ $\mathcal{U}_{9/2},$ $\mathcal{U}_{5},$ 
$\mathcal{U}_{11/2L},$ $\mathcal{U}_{6L},$ $\mathcal{U}_{13/2L}.$  Many additional eccentricity
functions attached to individual powers of $a$ were also found in closed form.  Lastly, we restate the curious connection 
between the redshift terms $\mathcal{U}_{11/2L}, \mathcal{U}_{6L},\mathcal{U}_{15/2}^{S2}$ and counterpart terms in the 
energy flux \cite{MunnETC23}.  The proportionality likely points to a common source of multipolar dependence, but the 
deeper significance of this connection will be left to future work. The full expansions 
can be found in the online repositories \cite{BHPTK18, UNCGrav22}.

It is likely that with a deeper expansion in eccentricity, more terms with rational coefficients throughout the redshift 
series could be manipulated into closed form.  However, for the purposes of transcribing these expansions into 
usable EOB models or waveform templates, the more important task lies in determining the multipolar dependence
of $\mathcal{U}_{5}^{\rm Sch}$ and then $\mathcal{U}_{11/2}^{S1}$, as these are the last components with unknown
contributions through 5.5PN order.  Once such an understanding is developed, the full eccentricity- and spin-dependence 
of the redshift invariant series for Kerr equatorial EMRIs will be known through 5.5PN order.   Note that the 
Schwarzschild limit of the redshift invariant was needed to 9.5PN and $e^8$ to complete a useful derivation of the 
scattering angle to 6PN within a framework that combines PN theory, PM theory, and EOB theory, implying
that these expansions continue to provide utility to high order \cite{BiniDamoGera20b}.  Nevertheless, the extent of 
the relationship between BHPT-PN expansions in the Kerr case and the EOB Hamiltonian is still an active area of 
study \cite{BiniGera19a}.  

By combining our these results with the extended Schwarzschild expansions of \cite{MunnEvan22a}, we were able
to make comparisons to numerical results in Fig.~\ref{fig:p10comps} and Fig.~\ref{fig:p20comps}, finding agreement to
better than $10^{-4}$ in most cases.  We also discovered interesting trends in the asymptotic behavior of the series across
values of $p, e,$ and $a$.  While the first two mostly followed expectations, the last showed a decreasing convergence with
$a$ that could point to a reduced efficacy of the expansion in the high-spin regime.  Further research into this question will be 
left to future work.

The techniques developed here can be utilized to expand the spin-precession invariant $\psi$ for Kerr 
equatorial EMRIs \cite{DolaETC14a, AkcaDempDola17, MunnEvan22b}.  This conservative quantity requires the 
expansion of the self-force, which involves first derivatives of the metric perturbations, along with the gauge portion 
of the metric completion piece \cite{BiniETC18, BiniGera19d}.  In the Schwarzschild case, the spin-precession
invariant incurred a factor of 5-10 greater computational expense, and the expansion process loses one order in $1/p$
and three orders in $e$ \cite{MunnEvan22b}.  Thus, the expectation is that the PN series there will be less extensive
than what we are able to get from the redshift.  Nevertheless, we should be able to extract some closed-form 
expressions, particularly at low orders, which will be fruitful as input for EOB models with spin.

Finally, with the PN behavior of the equatorial problem well understood, we will then be able to study the effects of 
inclination.  EMRI behavior is greatly complicated by the $\th$-motion, particularly in the source integration.  The
authors of \cite{IsoyETC22} were recently able to calculate the fluxes for generic EMRIs to 5PN/$e^{10}$.  It is 
expected that several of the computational simplifications applied here in the equatorial case will be applicable to 
generic orbits.  In particular, the MST homogeneous solutions take identical forms in both cases.  Thus, we may 
have the opportunity to extend those results.  The conservative sector will be more difficult still, as the metric 
perturbation expressions are significantly more cumbersome, and the $m$ summation formulas derived for 
spheroidal harmonics in Section~\ref{sec:genLexps} relied on simplifications in the equatorial plane.  Nevertheless,
the potential remains to derive yet undiscovered closed-form PN terms at low orders.  These
possibilities will all be explored in future work.

\acknowledgements

The author thanks Charles R. Evans, Jezreel Castillo, Scott Hughes, Zachary Nasipak, David Brown, Adrian Ottewill, 
Niels Warburton, Barry Wardell, and Chris Kavanagh for many helpful discussions in the preparation of this 
manuscript, and again thanks Zachary Nasipak for supplying the numerical redshift data.  This work was supported by NSF 
Grant Nos.~PHY-1806447 and PHY-2110335 to the University of North Carolina--Chapel Hill.  C.M.M.~acknowledges 
additional support from NASA ATP Grant 80NSSC18K1091 to MIT.  This work makes use of the black hole perturbation
toolkit.

\bibliography{redshiftK}

\end{document}